\documentclass[twocolumn, journal]{IEEEtran}
\IEEEoverridecommandlockouts

\usepackage{color}
\usepackage{xcolor}
\usepackage{bm}
\usepackage{graphicx}
\usepackage{amsmath}
\usepackage{amssymb}
\usepackage{multirow}
\usepackage{algorithm}
\usepackage{algorithmic}
\usepackage{booktabs}
\usepackage{array}
\usepackage{amsthm}
\usepackage{lipsum}
\usepackage{enumerate}
\usepackage{stfloats}
\usepackage{subfigure}
\usepackage{cases}
\usepackage{cite}
\usepackage{mathtools}
\usepackage{epstopdf}

\def\C{\mathbb{C}}

\def\Y{\mathbf{Y}}

\allowdisplaybreaks

\newtheorem{proposition}{Proposition}

\newtheoremstyle{noparens}%
  {}{}%
  {\itshape}{}%
  {\bfseries}{.}%
  { }%
  {\thmname{#1}\thmnumber{ #2}\mdseries\thmnote{ #3}}

\theoremstyle{noparens}

\title{Joint Training Scattering Matrix Learning and Channel Estimation for Beyond-Diagonal Reconfigurable Intelligent Surfaces}

\author{
    Yiyang Peng, Binggui Zhou,~\IEEEmembership{Member, IEEE}, Yutong Zheng, Danilo P. Mandic,~\IEEEmembership{Fellow, IEEE}, and Bruno Clerckx,~\IEEEmembership{Fellow, IEEE}
\thanks{This work has been partially supported by the UK Research and Innovation (UKRI) under Grant EP/Y004086/1, Grant EP/X040569/1, Grant EP/Y037197/1, Grant EP/X04047X/1, and Grant EP/Y037243/1.  \textit{(Corresponding author: Bruno Clerckx.)}}
\thanks{
Yiyang Peng, Yutong Zheng and Danilo Mandic are with the Department of Electrical and Electronic Engineering, Imperial College London, London SW7 2AZ, U.K. (e-mail: ~\{yiyang.peng22;yutong.zheng120;d.mandic\}@imperial.ac.uk).}
\thanks{Binggui Zhou is with the Department of Computer Science, Delft University of Technology, Delft 2628 CD, The Netherlands (e-mail: bingguizhou@tudelft.nl).}
\thanks{Bruno Clerckx is with the Department of Electrical and Electronic Engineering, Imperial College London, London SW7 2AZ, U.K., and also with the Department of Electronic Engineering, Kyung Hee University, Yongin-si, Gyeonggi-do 17104, South Korea (e-mail:~b.clerckx@imperial.ac.uk).}}

\begin{document}
\maketitle

\begin{abstract}
Beyond-diagonal reconfigurable intelligent surface (BD-RIS) generalizes the conventional diagonal RIS (D-RIS) by introducing tunable inter-element connections, offering enhanced wave manipulation capabilities. However, realizing the advantages of BD-RIS requires accurate channel state information (CSI), whose acquisition becomes significantly more challenging due to the increased number of channel coefficients, leading to prohibitively large pilot training overhead in BD-RIS-aided multi-user multiple-input multiple-output (MU-MIMO) systems. Existing studies reduce pilot overhead by exploiting the channel correlations induced by the Kronecker-product or multi-linear structure of BD-RIS-aided channels, which neglect the spatial correlation among antennas and the statistical correlation across RIS-user channels. In this paper, we propose a learning-based channel estimation framework, namely the joint training scattering matrix learning and channel estimation framework (JTSMLCEF), which jointly optimizes the BD-RIS training scattering matrix and estimates the cascaded channels in an end-to-end manner to achieve accurate channel estimation and reduce the pilot overhead. The proposed JTSMLCEF follows a two-phase channel estimation protocol to enable adaptive training scattering matrix optimization with a training scattering matrix optimizer (TSMO) and cascaded channel estimation with a dual-attention channel estimator (DACE). Specifically, the DACE is designed with intra-user and inter-user attention modules to capture the multi-dimensional correlations in multi-user cascaded channels. Simulation results demonstrate the superiority of JTSMLCEF. Compared with the current state-of-the-art method, it reduces the pilot overhead by $80\%$ while further reducing the normalized mean squared error (NMSE) by $82.6\%$ and $92.5\%$ in indoor and urban micro-cell (UMi) scenarios, respectively.
\end{abstract}

\begin{IEEEkeywords}
Beyond-diagonal reconfigurable intelligent surface (BD-RIS), low-overhead channel estimation, training scattering matrix learning, deep learning.
\end{IEEEkeywords}

\section{Introduction}
Beyond-diagonal reconfigurable intelligent surface (BD-RIS) represents an advanced RIS technique, which generalizes and extends conventional diagonal RIS (D-RIS) \cite{smart_radio_Marco_Di_renzo,IRS_tutorial_Qingqing_Wu,overview_signal_processing_ris_Pan}, by enabling reconfigurable inter-element connections among RIS elements via a multiport reconfigurable impedance network \cite{BD-RIS_magazine,bd_ris_survey_hongyu_ieee}. Compared to D-RIS, tunable inter-element connections introduce new degrees of freedom for BD-RIS, enabling smarter wave manipulation, broader coverage, and better performance across a variety of communication performance metrics and wireless applications \cite{BD-RIS_magazine,bd_ris_survey_hongyu_ieee}.

Motivated by these advantages, extensive research has been conducted to explore the benefits of BD-RIS from multiple perspectives, including architecture design \cite{shen2021modelling,nerinigraphtheory,band_stem_zheyu,architecture_discovery_zhou,hybrid_hongyu,multisector_hongyu}, optimization and performance analysis \cite{close_form_nerini,universal_optimization_zheyu,optimization_bd_ris_unified_approach}, and hardware impairments \cite{discrete_nerini,mutual_coupling_hongyu,lossy_interconnections_nerini,wideband_circuit_hongyu,lossy_bd_ris_peng_ieee,general_model_zheyu}. However, it should be noted that the performance gains offered by BD-RIS critically depend on accurate channel state information (CSI). For D-RIS, both pilot-based channel estimation methods \cite{survey_ce_irs_zheng,Swindlehurst_channel_estimation_ris_general_framework} and power-measurement-based methods \cite{sun_irs_reflection_design,sun_power_mensurement_ris_ce_wideband,sun_power_measurement_ris_ce} have been extensively investigated. Unlike D-RIS, the circuit topology of BD-RIS requires part of/all scattering elements to be jointly tuned rather than independently tuned. As a result, existing cascaded channel estimation methods developed for D-RIS cannot be directly applied to BD-RIS. Moreover, the inter-element connections of BD-RIS significantly increase the number of effective channel coefficients, leading to substantially higher training overhead.

The motivation of this work is twofold. First, base station (BS)-RIS-user cascaded channels in BD-RIS-aided multi-user multiple-input multiple-output (MU-MIMO) systems exhibit strong intra-user and inter-user correlations. Specifically, intra-user correlations arise from the common BS-RIS channel in the Kronecker-product structure of the cascaded channel and from spatial dependencies among transmit and receive antennas, while inter-user correlations arise from the statistical correlations among RIS-user channels across different users and the shared BS-RIS channel. Exploiting these correlations can significantly reduce the number of independent channel coefficients, thereby reducing the pilot overhead. Besides, these correlations induce rich feature patterns in the cascaded channels that can be effectively learned by deep learning-based methods to enhance channel estimation accuracy. Although the channel estimation design in \cite{channel_estimation_hongyu} is optimal in terms of minimizing mean squared error (MSE) under the least square (LS) estimator, it treats the cascaded channel coefficients as independent unknowns, without considering the correlations of cascaded channels, inducing a
large pilot overhead and limited practical applicability. Several model-driven approaches have been proposed to reduce the training overhead by exploiting channel correlations \cite{low_overhead_ce_bd_ris,lowoverhead_bd_ris_group,tensor_estiamte_ce_bd_ris,tensor_predict_estimate_ce_bd_ris,semi_blind_tensor_ce_bd_ris}, while a recent power-measurement-based method estimates BD-RIS channels from received-power observations \cite{power_measurement_ce_bd_ris}. In \cite{low_overhead_ce_bd_ris}, an explicit channel correlation model induced by the Kronecker-product structure is exploited to develop a two-phase channel estimation method for fully-connected BD-RIS. More recently, \cite{lowoverhead_bd_ris_group} extends this approach to group-connected BD-RIS, which estimates the reference cascaded channels of all groups using common received pilot signals in Phase I and subsequently estimates the remaining scaling coefficients in Phase II, thereby avoiding redundant group-by-group reference-channel training and further reducing the pilot overhead. Tensor-based methods \cite{tensor_estiamte_ce_bd_ris,tensor_predict_estimate_ce_bd_ris,semi_blind_tensor_ce_bd_ris} exploit the multi-linear structure of BD-RIS-aided channels through tensor decomposition. However, these model-driven approaches exhibit several limitations. In particular, \cite{tensor_estiamte_ce_bd_ris,semi_blind_tensor_ce_bd_ris} consider only single-user systems, \cite{tensor_predict_estimate_ce_bd_ris} neglects inter-user correlations, and all of them \cite{low_overhead_ce_bd_ris,tensor_estiamte_ce_bd_ris,tensor_predict_estimate_ce_bd_ris,semi_blind_tensor_ce_bd_ris,lowoverhead_bd_ris_group} do not exploit the spatial correlations among transmit/receive antennas and statistical correlations among RIS-user channels across users. In addition, the methods in \cite{low_overhead_ce_bd_ris,lowoverhead_bd_ris_group} rely on a sequential estimation procedure, such that errors in the estimated reference channels propagate to the subsequent estimation of the scaling coefficients. Moreover, tensor decomposition methods \cite{tensor_estiamte_ce_bd_ris,tensor_predict_estimate_ce_bd_ris,semi_blind_tensor_ce_bd_ris} rely on iterative optimization as well as explicit rank or identifiability assumptions, leading to computational inefficiency and sensitivity to model mismatch in dynamic scattering environments.

By contrast, recent advances in deep learning-based channel estimation have demonstrated that deep neural networks (DNNs) can effectively learn complex channel characteristics by implicitly and adaptively exploiting the multi-dimensional correlations of wireless channels in massive MIMO systems \cite{zhou_dual_attention} and capturing inter-user interactions in D-RIS-aided systems \cite{learning_to_reflect_and_beamforming}, showing improved estimation accuracy and reduced pilot overhead. Additionally, learning-based channel estimation methods do not require iterative optimization after online deployment, which is computationally efficient in practice. As such, it is promising to investigate learning-based channel estimation to exploit the multi-dimensional channel correlations in BD-RIS-aided MU-MIMO systems for accurate channel estimation and reduced pilot overhead.

Second, the design of training patterns plays an important role in improving channel estimation accuracy for RIS-aided systems. Most pilot-based D-RIS channel estimation methods \cite{survey_ce_irs_zheng,Swindlehurst_channel_estimation_ris_general_framework} adopt orthogonality-based training patterns under general channel models \cite{learning_to_reflect_and_beamforming} or structured channel models \cite{gui_zhou_channel_estimation_ris} to minimize the channel estimation error. However, such training pattern designs for D-RIS cannot be directly extended to BD-RIS due to different physical constraints imposed by the reconfigurable impedance network. A few works have investigated training pattern design for BD-RIS channel estimation \cite{channel_estimation_hongyu,tensor_estiamte_ce_bd_ris,semi_blind_tensor_ce_bd_ris,low_overhead_ce_bd_ris}, all of which adopt unitarity-based designs developed under the assumption of a non-reciprocal impedance network. However, in practice, the reconfigurable impedance network is typically reciprocal, since realizing non-reciprocal networks requires embedding non-reciprocal circuits, which is often cost-inefficient and impractical in real deployments. Reciprocity further imposes an additional physical constraint on BD-RIS, which mathematically results in a symmetric scattering matrix. Consequently, existing unitarity-based training designs developed for non-reciprocal BD-RIS are no longer applicable. Moreover, the symmetry constraint restricts the feasible training pattern space, thereby complicating the training pattern design for reciprocal BD-RIS. As a result, the design of effective training patterns for channel estimation in reciprocal BD-RIS remains an open problem.

Based on these motivations, in this work, we propose a learning-based channel estimation framework for BD-RIS-aided MU-MIMO systems, namely the joint training scattering matrix learning and channel estimation framework (JTSMLCEF), to jointly optimize the BD-RIS training scattering matrix and estimate the cascaded channels in an end-to-end manner, in order to achieve accurate channel estimation and reduce the pilot overhead. Unlike existing model-driven approaches in \cite{low_overhead_ce_bd_ris,tensor_estiamte_ce_bd_ris,tensor_predict_estimate_ce_bd_ris,semi_blind_tensor_ce_bd_ris}, the proposed JTSMLCEF exploits multi-dimensional channel correlations without estimation error propagation or relying on real-time iterative optimization procedures. The main contributions of this work are summarized in the following.
\begin{itemize}
    \item We exploit the structure of effective channel under reciprocal group-connected BD-RIS to reformulate the cascaded channel by combining channel coefficients that share the same scattering coefficients. This leads to an equivalent reduced-coefficient reformulation of the cascaded channel, significantly reducing the number of channel coefficients that need to be estimated. In contrast, existing BD-RIS channel estimation works either assume non-reciprocal BD-RIS architectures or neglect this structural property of the effective channel.
    \item The proposed JTSMLCEF follows a two-phase channel estimation protocol tailored for time division duplexing (TDD) BD-RIS-aided systems, enabling adaptive training scattering matrix design and uplink channel estimation within each channel coherence block. Specifically, the uplink training process consists of Phase I for optimizing the scattering matrix from received pilot observations with a training scattering matrix optimizer (TSMO), and Phase II for cascaded channel estimation with a dual-attention channel estimator (DACE) using the optimized scattering matrices. Unlike most existing D-RIS and BD-RIS training protocols that devote all training slots to channel estimation with fixed channel-independent training patterns, the proposed two-phase protocol allocates a small portion of the training slots to optimize the scattering matrices first. This enables a channel-dependent training pattern design that is dynamically adapted to instantaneous channel conditions, thereby significantly improving estimation efficiency.
    \item To minimize the estimation error and address the challenge of incorporating unitary and symmetric constraints of the scattering matrix, we design the TSMO to optimize the training susceptance matrix and convert it into a feasible training scattering matrix. Meanwhile, to exploit both intra-user and inter-user correlations inherent in multi-user cascaded BD-RIS channels, we propose the DACE architecture with two novel intra- and inter-user attention modules based on the multi-head self-attention (MHSA) mechanism \cite{attention_is_all_you_need}. By jointly learning the training scattering matrix via the TSMO and the intra- and inter-user dimension features via the DACE in an end-to-end fashion, the proposed JTSMLCEF achieves high estimation accuracy with significantly low pilot overhead.
    \item We present simulation results to verify the effectiveness of the proposed JTSMLCEF using QuaDRiGa channel model. In both indoor and urban micro-cell (UMi) scenarios, JTSMLCEF consistently achieves the lowest normalized mean squared error (NMSE) across various uplink transmit power levels, number of RIS elements, and training time slots, compared with the LS method, model-driven approach in \cite{low_overhead_ce_bd_ris} and deep-learning baselines. The ablation results validate the contributions of TSMO and DACE, while the dynamic-user experiments demonstrate the adaptability of a single trained model to different numbers of scheduled users.
\end{itemize}

\textit{Organization:}
Section~\ref{sec:system_model} introduces the system model. Section~\ref{sec:two_phase_protocol} presents the proposed two-phase channel estimation protocol. In Section~\ref{sec:J_TSML_CE_Network}, we introduce the training scattering matrix learning and dual-attention channel estimation in detail. Section~\ref{sec:simulations} provides simulation results that verify the effectiveness of JTSMLCEF and compare its performance with baseline methods. Finally, Section~\ref{sec:conclusion} concludes this work.

\textit{Notations:}
Non-bold letters, boldface lowercase letters, boldface uppercase letters, and underlined boldface uppercase letters represent scalars, column vectors, matrices, and tensors, respectively. $[\mathbf{a}]_i$ and $[\mathbf{A}]_{i,j}$ denote the $i$-th entry of $\mathbf{a}$ and the $(i,j)$-th entry of $\mathbf{A}$, respectively. $[\mathbf{A}]_{i:i',\,j:j'}$ represents a submatrix of $\mathbf{A}$ formed by extracting the $i$-th to $i'$-th rows and $j$-th to $j'$-th columns. Similarly, $[\underline{\mathbf{A}}]_{i:i',\,j:j',\,k:k'}$ denotes a sub-tensor of $\underline{\mathbf{A}}$ formed by the specified index ranges along the first, second, and third dimensions, respectively. This notation naturally extends to higher-order tensors. $\mathbb{R}$ and $\mathbb{C}$ denote the set of real numbers and complex numbers, respectively. $\Re\{\cdot\}$ and $\Im\{\cdot\}$ take the real and imaginary parts of the input, respectively. $(\cdot)^T$, $(\cdot)^H$, and $(\cdot)^{-1}$ represent transpose, conjugate transpose, and inverse operations, respectively. $\otimes$ and $\odot$ represent the Kronecker product and Hadamard product, respectively. $|\cdot|$, $\| \cdot \|_2$ and $\| \cdot \|_F$ denote the absolute-value norm, the $\ell_2$ norm, and the Frobenius norm, respectively. $\mathsf{vec}(\cdot)$ is the vectorization of a matrix, and $\overline{\mathsf{vec}}(\cdot)$ denotes the reverse operation of vectorization. $\mathbf{I}_M$ denotes an $M \times M$ identity matrix. $\mathsf{Tr}(\cdot)$ returns the trace of a matrix. Finally, $\jmath = \sqrt{-1}$ represents the imaginary unit.

\section{System Model}\label{sec:system_model}
We consider a narrowband BD-RIS-aided MU-MIMO system working in the TDD mode. As shown in Fig. \ref{fig:system_model}, the system consists of a BS equipped with $N$ antennas, an $M$-element BD-RIS, and $K$ users, each equipped with $U$ antennas. The $M$-element BD-RIS is a passive device modeled as $M$ scattering elements connected to an $M$-port reconfigurable impedance network \cite{shen2021modelling}. This reconfigurable network comprises a number of tunable components and can be mathematically characterized by its scattering matrix. The BS controls the BD-RIS through an RIS controller that adjusts the tunable components of the reconfigurable impedance network to steer the reflected signals toward desired directions.

\subsection{Channel Model}
We assume a quasi-static block-fading channel model in which the channel remains approximately constant within each coherence block but varies from block to block. During the uplink training phase, pilot signals are transmitted, and the resulting channel estimates are subsequently used for downlink data transmission. In the uplink pilot transmission, the effective channel $\mathbf{H}_{\mathrm{eff},k}\in\C^{N\times U}$ between each user $k$ and the BS is expressed as\footnote{We adopt commonly used channel modeling assumptions between the BS, BD-RIS, and users, including perfect antenna matching, no mutual coupling, unilateral approximation, and no structural scattering \cite{universal_framework_nerini}.}
\begin{equation}
    \mathbf{H}_{\mathrm{eff},k} = \mathbf{H}_{RT,k} + \mathbf{H}_{IT}\bm\Phi\mathbf{H}_{RI,k}, ~\forall k \in \mathcal{K} = \{1, 2,\ldots, K\},\label{eq:H_eff}
\end{equation}
where $\mathbf{H}_{RT,k}\in\C^{N \times U}$, $\mathbf{H}_{RI,k}\in\C^{M \times U}$ and $\mathbf{H}_{IT}\in\C^{N \times M}$ denote the direct user-BS, user-RIS, and RIS-BS channels for user~$k$, respectively, and $\bm\Phi\in\C^{M\times M}$ is the uplink scattering matrix. In this work, we assume that the direct user-BS channels $\mathbf{H}_{RT,k},~\forall k\in\mathcal{K}$, are blocked and focus on estimating the cascaded user-RIS-BS channels\footnote{When the direct user-BS channels exist, they can be estimated by first turning off the RIS and applying conventional MU-MIMO channel estimation methods; their contributions can then be subtracted from the received signals.}. The uplink effective channel between user $k$ and BS is then given by 
\begin{equation}
    \mathbf{H}_{\mathrm{eff},k} = \mathbf{H}_{IT}\bm\Phi\mathbf{H}_{RI,k} 
    = \overline{\mathsf{vec}}\left(\mathbf{Q}_k ~ \mathsf{vec}(\bm\Phi)\right),\label{eq:H_eff_Q_k_uplink}
\end{equation}
where $\mathbf{Q}_k=\mathbf{H}_{RI,k}^T \otimes \mathbf{H}_{IT}\in\C^{NU\times M^2}$ denotes the uplink cascaded user-RIS-BS channel for user $k$. In \cite{channel_estimation_hongyu}, it has been shown that in TDD systems, the uplink cascaded channel $\mathbf{Q}_k,~\forall k\in\mathcal{K}$, estimated from uplink training, can be directly used for downlink data transmission by utilizing the uplink-downlink reciprocity of $\mathbf{H}_{IT}$ and $\mathbf{H}_{RI,k},~\forall k\in\mathcal{K}$.

\begin{figure}
    \centering
    \includegraphics[width=0.43\textwidth]{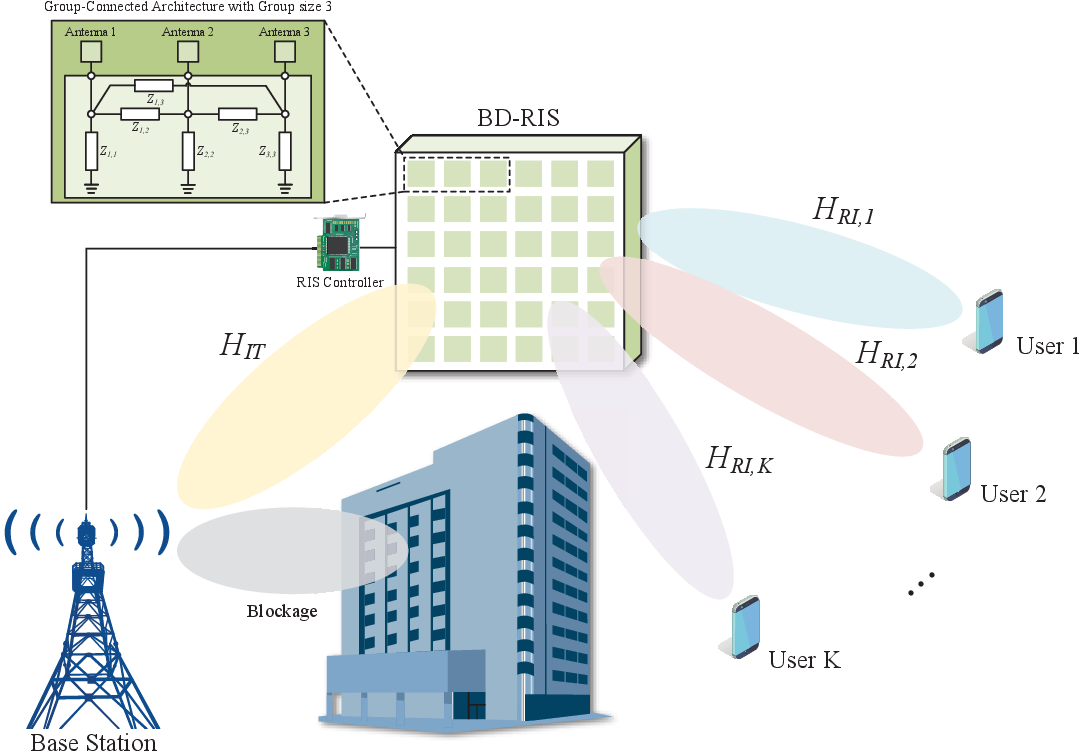}
    \caption{A BD-RIS-aided MU-MIMO communication system.}\label{fig:system_model}
\end{figure}

Depending on the circuit topology of the $M$-port reconfigurable impedance network, various BD-RIS architectures have been proposed in the literature \cite{shen2021modelling,nerinigraphtheory,band_stem_zheyu}. In this work, we focus on the group-connected architecture of BD-RIS \cite{shen2021modelling}. For the group-connected BD-RIS, the $M$-port reconfigurable impedance network is uniformly divided into $G$ groups, with each group containing $\bar{M}=\frac{M}{G}$ elements, referred to as the group size. Fig.~\ref{fig:system_model} illustrates a 36-element BD-RIS with a group-connected architecture of group size 3, where one group is highlighted for illustration. Physically, the ports within the same group are mutually connected via
tunable components, e.g., $Z_{1,2},Z_{1,3}$ and $Z_{2,3}$ in Fig.~\ref{fig:system_model}, whereas ports belonging to different groups are not connected. Mathematically, a group-connected BD-RIS exhibits a block-diagonal scattering matrix $\bm{\Phi}$, written as
\begin{equation}
    \bm\Phi = \mathsf{blkdiag}(\bm\Phi_1,\bm\Phi_2,\ldots,\bm\Phi_G),\label{eq:blkdiag}
\end{equation}
where each block $\bm\Phi_g\in\C^{\bar{M}\times\bar{M}},~\forall g\in\mathcal{G}=\{1,2,\ldots,G\},$ represents the scattering behavior of group $g$. In this work, we assume that the reconfigurable impedance network is lossless and reciprocal, which leads to each block $\bm{\Phi}_g$ being unitary and symmetric, respectively, i.e.,
\begin{equation}
    \bm\Phi_g^H \bm\Phi_g = \mathbf{I}_{\bar{M}},~ \bm\Phi_g^T=\bm\Phi_g,~\forall g\in\mathcal{G}.\label{eq:Phi_unitary_symmetric}
\end{equation}

\subsection{A Reduced-Coefficient Reformulation of the Cascaded Channel}
Under the group-connected BD-RIS architecture, the effective channel in \eqref{eq:H_eff_Q_k_uplink} can be equivalently rewritten as
\begin{equation}
    \mathbf{H}_{\mathrm{eff},k}=\sum_{g\in\mathcal{G}} \mathbf{H}_{IT,g}\bm\Phi_{g}\mathbf{H}_{RI,k,g}=\overline{\mathsf{vec}}\Big(\sum_{g\in\mathcal{G}}\tilde{\mathbf{Q}}_{k,g}\bm\phi_g\Big),\label{eq:H_eff_Q_tilde_phi}
\end{equation}
where $\mathbf{H}_{IT,g}=[\mathbf{H}_{IT}]_{:,(g-1)\bar{M}+1:g\bar{M}}\in\C^{N\times \bar{M}}$, $\mathbf{H}_{RI,k,g}=[\mathbf{H}_{RI,k}]_{(g-1)\bar{M}+1:g\bar{M},:}\in\C^{\bar{M}\times U}$, denoting the $g$-th block of $\mathbf{H}_{IT}$ and $\mathbf{H}_{RI,k}$, respectively, $\bm\phi_{g}\triangleq\mathsf{vec}(\bm\Phi_{g})\in\C^{\bar{M}^2\times 1},~\forall g\in\mathcal{G}$, denoting the $g$-th block of vectorized BD-RIS scattering matrix, and \begin{equation}
    \tilde{\mathbf{Q}}_{k,g}\triangleq \mathbf{H}_{RI,k,g}^T \otimes \mathbf{H}_{IT,g} \in\C^{NU\times \bar{M}^2},~\forall g\in\mathcal{G},~\forall k\in\mathcal{K},\label{eq:tilde_Q_k_g}
\end{equation}
denoting the $g$-th group of the uplink cascaded channel for user $k$. For a reciprocal BD-RIS, the scattering matrix $\bm\Phi$ is symmetric. This property leads to the following reduced-coefficient reformulation.

\begin{proposition}\label{proposition:reduced_coefficient}
For a reciprocal group-connected BD-RIS, i.e., $\bm\Phi_g^T=\bm\Phi_g,~\forall g\in\mathcal{G}$, the effective channel in \eqref{eq:H_eff_Q_tilde_phi} can be equivalently expressed as
\begin{equation}
\mathbf{H}_{\mathrm{eff},k}=\overline{\mathsf{vec}}\left(\bar{\mathbf{Q}}_{k}\bar{\bm\phi}\right),\label{eq:H_eff_Q_bar_phi}
\end{equation}
where $\bar{\bm\phi}=\big[\bar{\bm\phi}_{1}^T,\bar{\bm\phi}_{2}^T,\ldots,\bar{\bm\phi}_{G}^T\big]^T\in\C^{\frac{M(\bar{M}+1)}{2}\times 1}$, collects the unique scattering coefficients, e.g., the diagonal and
upper-triangular entries of $\{\bm\Phi_g\}_{g=1}^G$, and \begin{equation}
    \bar{\mathbf{Q}}_{k}=\big[\bar{\mathbf{Q}}_{k,1},\bar{\mathbf{Q}}_{k,2},\ldots,\bar{\mathbf{Q}}_{k,G}\big]\in\C^{NU\times \frac{M(\bar{M}+1)}{2}},\label{eq:bar_Q_k}
\end{equation}
is the reduced-coefficient cascaded channel of user $k$, with \begin{equation}
    \bar{\mathbf{Q}}_{k,g}\triangleq \tilde{\mathbf{Q}}_{k,g}\mathbf{P}\in\C^{NU\times \frac{\bar{M}(\bar{M}+1)}{2}},~\forall g\in\mathcal{G},~\forall k\in\mathcal{K}.\label{eq:bar_Q_k_g}
\end{equation}
Here, $\mathbf{P} \in \{0,1\}^{\bar{M}^2 \times \frac{\bar{M}(\bar{M}+1)}{2}}$ is a binary mapping matrix satisfying $\bm\phi_{g}=\mathbf{P}\bar{\bm\phi}_{g},~\forall g\in\mathcal{G}$, where $\bar{\bm\phi}_g\in \C^{\frac{\bar{M}(\bar{M}+1)}{2} \times 1}$ collects the diagonal and upper-triangular entries of $\bm\Phi_g$.
\end{proposition}

\begin{proof} 
The symmetry of $\bm\Phi_g,~\forall g\in\mathcal{G}$, implies that every lower-triangular entry equals its corresponding upper-triangular entry. Hence, its vectorization satisfies $\bm\phi_g=\mathbf P\bar{\bm\phi}_g$, where the explicit form of $\mathbf{P}$ is given in the Appendix of \cite{lossy_bd_ris_peng_ieee}. Substituting this relation into \eqref{eq:H_eff_Q_tilde_phi} gives $\sum_{g\in\mathcal G} \tilde{\mathbf Q}_{k,g}\bm\phi_g = \sum_{g\in\mathcal G} \tilde{\mathbf Q}_{k,g}\mathbf P\bar{\bm\phi}_g = \bar{\mathbf Q}_k\bar{\bm\phi}, $ which yields \eqref{eq:H_eff_Q_bar_phi}. Since this equality holds for every reciprocal scattering matrix, the reformulation is exact and introduces no approximation. 
\end{proof}

To illustrate the reduced-coefficient reformulation in Proposition~\ref{proposition:reduced_coefficient}, consider the case $\bar M=2$, for which 
\begin{equation}
\bm\Phi_g= \begin{bmatrix} \phi_1&\phi_2\\ \phi_2&\phi_3 \end{bmatrix},~\bar{\bm\phi}_g=\begin{bmatrix} \phi_1\\\phi_2\\\phi_3 \end{bmatrix},~\mathbf P= \begin{bmatrix} 1&0&0\\ 0&1&0\\ 0&1&0\\ 0&0&1 \end{bmatrix}.
\end{equation}
Let $\tilde{\mathbf Q}_{k,g}=[\mathbf q_{11}~\mathbf q_{21}~\mathbf q_{12}~\mathbf q_{22}]$. Then $ \tilde{\mathbf Q}_{k,g}\bm\phi_g = \mathbf q_{11}\phi_1 + \left(\mathbf q_{21}+\mathbf q_{12}\right)\phi_2 + \mathbf q_{22}\phi_3$. Thus, $\mathbf q_{21}$ and $\mathbf q_{12}$ enter every observation only through their sum and cannot be identified individually. The reduced-coefficient matrix is therefore $\bar{\mathbf Q}_{k,g} = \tilde{\mathbf Q}_{k,g}\mathbf P = [\mathbf q_{11}~(\mathbf q_{21}+\mathbf q_{12})~\mathbf q_{22}]$. Consequently, the multiplication $\tilde{\mathbf Q}_{k,g}\mathbf P$ combines the columns associated with the same scattering coefficient while preserving exactly all coefficient combinations that can affect the effective channel.

With reduced-coefficient reformulation $\bar{\mathbf{Q}}_{k}$ for each user $k$, the total number of channel coefficients to be estimated per user reduces from $NUM^2$ to $NU\frac{M(\bar{M}+1)}{2}$.\footnote{For a non-reciprocal BD-RIS, the scattering matrix is generally not symmetric, so there are generally no repeated scattering coefficients. Hence, the reduced-coefficient reformulation of the cascaded channels in \eqref{eq:bar_Q_k_g} does not apply.} Nevertheless, even after exploiting the symmetry and block-diagonal structures of the scattering matrix, the number of coefficients to be estimated in BD-RIS-aided channels remains prohibitively large. From \eqref{eq:tilde_Q_k_g}, \eqref{eq:bar_Q_k}, and \eqref{eq:bar_Q_k_g}, it can be observed that strong structural dependencies exist both within each $\bar{\mathbf{Q}}_{k}$ and across $\bar{\mathbf{Q}}_{k}$ of different users, owing to their block Kronecker structures. Specifically, each $\bar{\mathbf{Q}}_{k,g}$ is formed by the combination of $\mathbf{H}_{RI,k,g}$ and $\mathbf{H}_{IT,g}$, where all antennas of $\mathbf{H}_{RI,k,g}$ share the same $\mathbf{H}_{IT,g},~\forall g \in \mathcal{G}$. As a result, the channel coefficients in each $\bar{\mathbf{Q}}_{k}$ exhibit strong dependencies along both rows and columns, as well as across different users. Moreover, the multiplication by the binary matrix $\mathbf{P}$ in \eqref{eq:bar_Q_k_g} preserves these structural dependencies. These observations motivate us to design a learning-based channel estimation scheme that can effectively exploit these correlations to achieve accurate channel estimation with significantly reduced pilot overhead.

\textit{Remark 1:} The reduced-coefficient reformulation in \eqref{eq:bar_Q_k_g} is applicable to any BD-RIS architectures whose scattering matrix is symmetric and block-diagonal\footnote{For reciprocal band-connected and stem-connected BD-RISs \cite{band_stem_zheyu,Q-stem_xiaohua}, the scattering matrix is not block-diagonal. In such cases, a different form of the mapping matrix $\mathbf{P}$ can be employed to extract unique scattering coefficients and achieve coefficient reduction.}. For example, in forest-connected \cite{nerinigraphtheory} and cluster-connected architectures \cite{general_architectures_xiaohua}, each $\bm{\Phi}_g$ is still a full symmetric matrix, so the same mapping matrix $\mathbf{P}$ can be used.

\section{Two-Phase Channel Estimation Protocol}\label{sec:two_phase_protocol}
In this section, we describe the pilot transmission procedure and the two-phase channel estimation protocol tailored to the proposed JTSMLCEF.

\subsection{Overall Description of the Two-Phase Protocol}
\begin{figure*}
    \centering
    \includegraphics[width=0.9\textwidth]{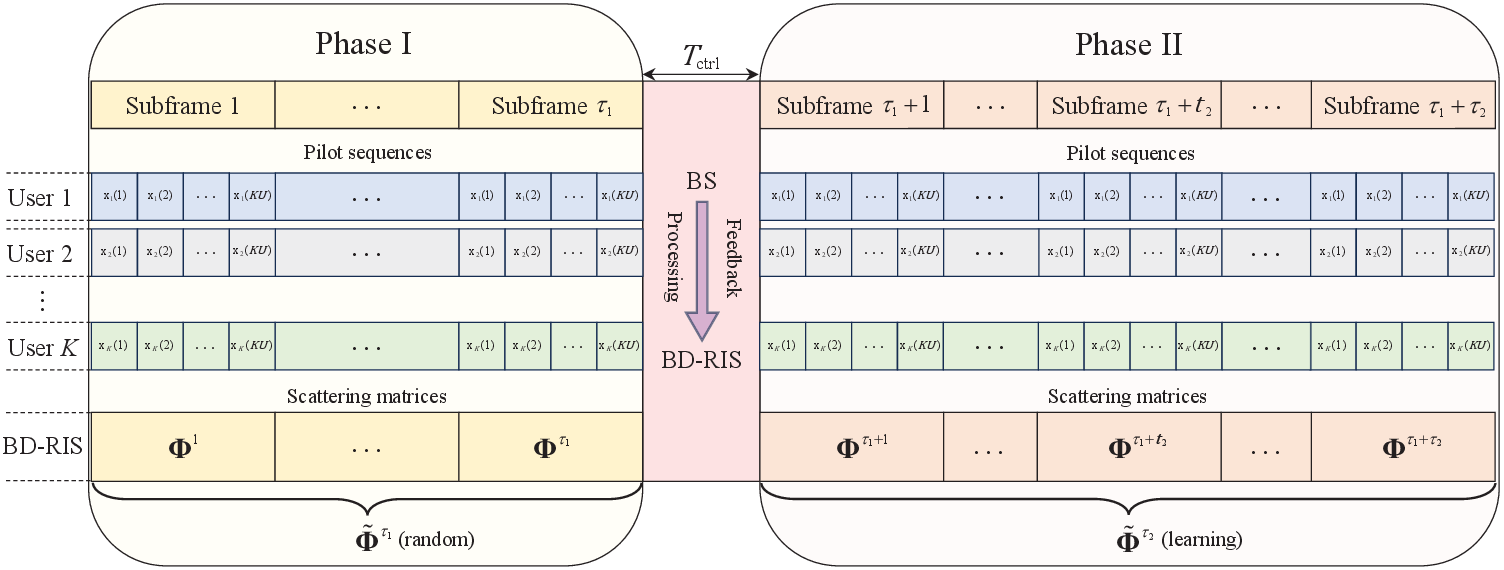}
    \caption{The overall two-phase channel estimation protocol. The total number of training time slots is $KU(\tau_1 + \tau_2)$. $T_{\mathsf{ctrl}}$ refers to the latency caused by BS-side processing, as well as the control and feedback signaling to the BD-RIS through the RIS controller.}\label{fig:channel_estimation_protocol}
\end{figure*}

To ensure that the designed training pattern effectively assists channel estimation, both the scattering matrix design and cascaded channels estimation are performed within the same coherence block. Since the ultimate objective is to obtain an accurate channel estimate from the received pilot signals at the BS, the training pattern is designed based on previously received pilot signals and then applied for subsequent pilot transmissions to minimize estimation error. Motivated by this principle, the uplink transmission within each coherence block is divided into two phases. In Phase I, the scattering matrices are optimized from the received pilot observations. In Phase II, cascaded channels are estimated based on the received pilots employing optimized scattering matrices. The overall two-phase channel estimation protocol is illustrated in Fig. \ref{fig:channel_estimation_protocol}.

\subsection{Pilot Transmission}
We assume that a total of $L$ time slots within each coherence block are allocated for uplink pilot transmission. At each time slot $l,~\forall l \in \mathcal{L}=\{1,2,\ldots,L\},$ user $k$ transmits a pilot sequence
\begin{equation}
    \mathbf{x}_k(l)=\big[[\mathbf{x}_{k}(l)]_1,[\mathbf{x}_{k}(l)]_2,\ldots,[\mathbf{x}_{k}(l)]_U\big]^T\in\mathbb{C}^{U\times 1},
\end{equation}
where $[\mathbf{x}_k(l)]_u\in\C$ denotes the symbol transmitted from the $u$-th antenna of user $k$. Each pilot symbol satisfies 
\begin{equation}
    \big|[\mathbf{x}_k(l)]_u\big| = 1,~\forall u\in\mathcal{U}=\{1, 2,\ldots, U\},~\forall k\in\mathcal{K}.
\end{equation}
The pilot signals propagate through the BD-RIS and are received at the BS. The received signal at the BS during time slot $l$ is given by
\begin{align}
    \mathbf{y}(l)&=\sqrt{P_u}\sum_{k=1}^K \mathbf{H}_{IT}\bm\Phi(l)\mathbf{H}_{RI,k}\mathbf{x}_k(l)+\mathbf{n}(l),\notag \\
    &=\sqrt{P_u}\mathbf{H}_{IT}\bm\Phi(l)\mathbf{H}_{RI}\mathbf{x}(l)+\mathbf{n}(l),\label{eq:received_pilots_y_l}
\end{align}
where $P_u$ denotes the per-antenna uplink transmit power of each user\footnote{We assume equal transmit power allocation across all user antennas.}, $\bm{\Phi}(l)$ is the BD-RIS scattering matrix applied at time slot $l$, $\mathbf{H}_{RI}=[\mathbf{H}_{RI,1},\mathbf{H}_{RI,2},\ldots,\mathbf{H}_{RI,K}]\in\C^{M\times KU}$ collects the user-RIS channels of all users, $\mathbf{x}(l)=\big[\mathbf{x}_1^T(l),\mathbf{x}_2^T(l),\ldots,\mathbf{x}_K^T(l)\big]^T\in\C^{KU\times 1}$ is the transmit pilot sequence of all users at time slot $l$, and $\mathbf{n}(l)\sim \mathcal{CN}(\mathbf{0}, \sigma^2 \mathbf{I}_N)\in\mathbb{C}^{N\times 1}$ denotes additive white Gaussian noise (AWGN). We adopt the commonly used training strategy \cite{gui_zhou_channel_estimation_ris}, which divides the total training duration of $L$ time slots into $\tau$ subframes, each containing $KU$ time slots, i.e., $L = KU\tau$. Within each subframe, all $K$ users simultaneously transmit their pilot signals, repeated over $\tau$ subframes. The pilot signal transmitted by all users $\mathbf{X}\in\C^{KU\times KU}$ is given by
\begin{equation}
    \mathbf{X}=\big[\mathbf{x}(1),\mathbf{x}(2),\ldots,\mathbf{x}(KU)\big]=\big[\mathbf{X}_1^T,\mathbf{X}_2^T,\ldots,\mathbf{X}_K^T\big]^T,
\end{equation}
where $\mathbf{X}_k\in\C^{U\times KU}$ denotes the pilot matrix of user $k$, constructed as $\mathbf{X}_k=\big[\mathbf{x}_k(1),\mathbf{x}_k(2),\ldots,\mathbf{x}_k(KU)\big],~\forall k \in\mathcal{K}$. Meanwhile, the BD-RIS scattering matrix remains fixed within each subframe and changes over different subframes. Specifically, for subframe $t,~\forall t\in\mathcal{T}=\{1,2,\ldots,\tau\}$, the $KU$ time slots share the same uplink scattering matrix, i.e., $\bm\Phi((t-1)KU+1)=\cdots=\bm\Phi(tKU)=\bm\Phi^t$, where $\bm\Phi^t$ denotes the scattering matrix applied in subframe $t$. Then, the received pilot signal at the BS in subframe $t$ can be expressed as
\begin{align}
    \Y^t &= \big[\mathbf{y}((t-1)KU+1),\ldots,\mathbf{y}(tKU)\big],\notag \\ &=\sqrt{P_u}\mathbf{H}_{IT}\bm\Phi^t\mathbf{H}_{RI}\mathbf{X}+\mathbf{N}^t,
\end{align}
where $\mathbf{N}^t\in\C^{N\times KU}$ represents the noise matrix in subframe $t$, with each column following the distribution $\mathcal{CN}(\mathbf{0}, \sigma^2 \mathbf{I}_N)$. To ensure that the BS can distinctly observe the contribution of each user and each antenna, the pilot matrices are designed to maintain both inter-user and intra-user orthogonality. Specifically, the pilot matrix of each user $\mathbf{X}_k$ satisfies $\mathbf{X}_{k_1} \mathbf{X}_{k_2}^H =KU\mathbf{I}_U,~\text{if}~k_1 = k_2,$ and $\mathbf{X}_{k_1} \mathbf{X}_{k_2}^H =\mathbf{0}_U,~\text{if}~k_1 \neq k_2,~\forall\, k_1, k_2 \in \mathcal{K}$. This ensures that pilot sequences transmitted from different antennas of the same user are mutually orthogonal, and the pilots transmitted by different users are also mutually orthogonal. Thus, the contribution of user $k$ in subframe $t$, denoted by $\Y_{k}^t\in\C^{N\times U}$, can be decorrelated from $\Y^t$ as
\begin{equation}
    \Y_{k}^t = \frac{1}{KU}\Y^t \mathbf{X}_k^H
    =\sqrt{P_u} \mathbf{H}_{IT}\bm\Phi^t\mathbf{H}_{RI,k} + \tilde{\mathbf{N}}^t,\label{eq:Y_k_t}
\end{equation}
$\forall k\in\mathcal{K},~\forall t\in\mathcal{T}$, where $\tilde{\mathbf{N}}^t=\frac{1}{KU}\mathbf{N}^t\mathbf{X}_k^H$ denotes the noise after decorrelation and the user index $k$ is omitted for simplicity. This training design, combined with the inter-user and intra-user pilot orthogonality, enables the BS to separate the pilot contribution of each user antenna and obtain multiple distinct observations of the channel. Leveraging the reduced-coefficient reformulation of cascaded channel in \eqref{eq:H_eff_Q_bar_phi}, \eqref{eq:Y_k_t} can be equivalently rewritten as
\begin{equation}
    \Y_{k}^t=\sqrt{P_u}~\overline{\mathsf{vec}}\left(\bar{\mathbf{Q}}_{k}\bar{\bm\phi}^t\right)+ \tilde{\mathbf{N}}^t,\label{eq:Y_k_t_Q_bar_phi_bar}
\end{equation}
where $\bar{\bm\phi}^t\in\C^{\frac{M(\bar{M}+1)}{2}\times 1}$ denotes the vector containing all unique scattering coefficients in subframe $t,~\forall t\in\mathcal{T}$. By collecting $\Y_{k}^t,~\forall k\in\mathcal{K}$, over $\tau$ uplink training subframes at the BS, the received pilot signal of user $k$ across $\tau$ subframes can be stacked to form $\Y_k^{\tau}\in\C^{NU\times \tau}$, which is given by
\begin{align}
    \Y_k^{\tau}&=\big[\mathsf{vec}(\Y_{k}^1),\mathsf{vec}(\Y_{k}^2),\ldots,\mathsf{vec}(\Y_{k}^\tau)\big],\notag\\
    &= \sqrt{P_u}\bar{\mathbf{Q}}_k \tilde{\bm\Phi} + \widetilde{\mathbf{N}},~\forall k\in\mathcal{K},
\end{align}
where $\widetilde{\mathbf{N}}=\big[\tilde{\mathbf{n}}^1,\tilde{\mathbf{n}}^2,\ldots,\tilde{\mathbf{n}}^\tau\big]\in\C^{NU\times \tau}$ with $\tilde{\mathbf{n}}^t=\mathsf{vec}(\tilde{\mathbf{N}}^{t}),$ $\forall t\in\mathcal{T}$, and 
\begin{equation}
\tilde{\bm\Phi}=\big[\bar{\bm\phi}^1,\bar{\bm\phi}^2,\ldots,\bar{\bm\phi}^\tau\big]\in\C^{\frac{M(\bar{M}+1)}{2}\times \tau}.
\end{equation} We refer to the collection of BD-RIS scattering matrices applied over $\tau$ uplink training subframes, i.e., $\tilde{\bm\Phi}$, as the \emph{overall BD-RIS training scattering matrix}. The total $\tau$ training subframes are divided into $\tau_1$ subframes for Phase I, which are used to design the training scattering matrix, and $\tau_2$ subframes for Phase II, which employ the optimized scattering matrix to obtain the final channel estimate, where $\tau = \tau_1 + \tau_2$.

\subsection{Phase I: Training Scattering Matrix Learning}
In Phase I, a sequence of BD-RIS scattering matrices $\bm\Phi^{t_1}$ is applied over $\tau_1$ subframes, where $t_1 \in \mathcal{T}_1 = \{1, \ldots, \tau_1\}$. By collecting the diagonal and upper-triangular entries of all BD-RIS groups from each $\bm\Phi^{t_1}$ across $\tau_1$ subframes, we define $\tilde{\bm\Phi}^{\tau_1}=\big[\bar{\bm\phi}^1,\bar{\bm\phi}^2,\ldots,\bar{\bm\phi}^{\tau_1}\big]\in\C^{\frac{M(\bar{M}+1)}{2}\times \tau_1},$ as the \emph{Phase-I training scattering matrix}. At this stage, no prior CSI is available at the BS. Therefore, $\tilde{\bm\Phi}^{\tau_1}$ is initialized randomly, with each scattering matrix $\bm\Phi^{t_1}$ satisfying the unitary and symmetric constraints in \eqref{eq:Phi_unitary_symmetric}. Within each subframe $t_1$, each user $k,~\forall k \in \mathcal{K}$, transmits the pilot symbols $\mathbf{X}_k$ to the BS. The received pilots of user $k$ over $\tau_1$ subframes in Phase I are then given by
\begin{equation}
\mathbf{Y}^{\tau_1}_k
= \sqrt{P_u}\bar{\mathbf{Q}}_k\tilde{\bm{\Phi}}^{\tau_1} + \widetilde{\mathbf{N}}^{\tau_1}
\in \mathbb{C}^{NU\times \tau_1},~\forall k\in\mathcal{K},\label{eq:Y_k_tau1}
\end{equation}
where $\widetilde{\mathbf{N}}^{\tau_1}=\big[\tilde{\mathbf{n}}^1,\tilde{\mathbf{n}}^2,\ldots,\tilde{\mathbf{n}}^{\tau_1}\big]$. The overall received pilot matrix collecting the observations of all users is given by $\Y^{\tau_1}\in\mathbb{C}^{NU\times \tau_1 K}$, which is constructed as $\mathbf{Y}^{\tau_1}=\big[\mathbf{Y}^{\tau_1}_1,\mathbf{Y}^{\tau_1}_2,\ldots,\mathbf{Y}^{\tau_1}_K\big]$, and can be reshaped into a tensor form $\underline{\Y}^{\tau_1}\in\C^{NU\times K\times\tau_1}$. Given $\underline{\Y}^{\tau_1}$, the information of cascaded channels is implicitly encoded in the received pilot observations. This enables the design of a more effective training scattering matrix by extracting channel-relevant features from $\underline{\Y}^{\tau_1}$ to improve channel estimation accuracy. Under the proposed two-phase protocol, explicit channel estimation is performed in Phase II. In Phase II, a set of distinct scattering matrices $\bm\Phi^{t_2},~\forall t_2\in\mathcal{T}_2=\{1,2,\ldots,\tau_2\}$, is applied across $\tau_2$ subframes to generate $\tau_2$ different received pilot observations. The scattering matrices applied in Phase II are collected as\begin{equation}
\tilde{\bm\Phi}^{\tau_2}=\big[\bar{\bm\phi}^1,\bar{\bm\phi}^2,\ldots,\bar{\bm\phi}^{\tau_2}\big]\in\C^{\frac{M(\bar{M}+1)}{2}\times \tau_2},
\end{equation}
which we refer to as the \emph{Phase-II training scattering matrix}\footnote{The overall
training scattering matrix consists of the Phase-I and Phase-II training scattering matrices, i.e., $\tilde{\bm\Phi}=[\tilde{\bm\Phi}^{\tau_1},\tilde{\bm\Phi}^{\tau_2}]$. The Phase-I matrix $\tilde{\bm\Phi}^{\tau_1}$ is randomly generated, whereas the Phase-II matrix $\tilde{\bm\Phi}^{\tau_2}$ is the design variable to be optimized, or learned, to improve the channel estimation performance.}. Unless otherwise specified, in the remainder of this paper, the term \emph{training scattering matrix} refers to the Phase-II training scattering matrix $\tilde{\bm\Phi}^{\tau_2}$. We design the TSMO to optimize the training scattering matrix based on $\underline{\mathbf{Y}}^{\tau_1}$. After processing $\underline{\mathbf{Y}}^{\tau_1}$, the BS computes the optimized $\tilde{\bm\Phi}^{\tau_2}$ and sends the parameters to the RIS controller. This operation introduces a control delay $T_{\mathsf{ctrl}}$, which can be approximated as
\begin{equation}
T_{\mathsf{ctrl}}
=
T_{\mathsf{inf}}+T_{\mathsf{link}},
\end{equation}
where $T_{\mathsf{inf}}$ is the online inference latency at the BS and $T_{\mathsf{link}}$ is the BS-to-RIS control-link transmission delay. The delay of $T_{\mathsf{ctrl}}$ is assumed to be negligible compared to the total uplink training time, i.e., $T_{\mathsf{ctrl}} \ll KU(\tau_1 + \tau_2)T_{\mathsf{slot}}$, where $T_{\mathsf{slot}}$ is the duration of one slot. The impact of $T_{\mathsf{ctrl}}$ relative to the uplink training duration is quantified in Section~\ref{sec:control_latency}.

\subsection{Phase II: Cascaded Channels Estimation}
In Phase II, the optimized training scattering matrix $\tilde{\bm\Phi}^{\tau_2}$ is applied at the BD-RIS. Using the same pilot symbols $\mathbf{X}_k$ as in Phase I, each user $k$ transmits orthogonal pilots to the BS over $\tau_2$ subframes. The received pilot observations of user $k$ over these $\tau_2$ subframes are given by
\begin{equation}
\mathbf{Y}^{\tau_2}_k
= \sqrt{P_u}\bar{\mathbf{Q}}_k\tilde{\bm{\Phi}}^{\tau_2} + \widetilde{\mathbf{N}}^{\tau_2}
\in \mathbb{C}^{NU\times \tau_2},~\forall k\in\mathcal{K},\label{eq:Y_k_tau2}
\end{equation}
where $\widetilde{\mathbf{N}}^{\tau_2}=\big[\tilde{\mathbf{n}}^1,\tilde{\mathbf{n}}^2,\ldots,\tilde{\mathbf{n}}^{\tau_2}\big]$. The overall received pilot matrix collecting the observations of all users is denoted by $\Y^{\tau_2}\in\mathbb{C}^{NU\times \tau_2 K}$, and is constructed as $\mathbf{Y}^{\tau_2}
=\big[\mathbf{Y}^{\tau_2}_1,\mathbf{Y}^{\tau_2}_2,\ldots,\mathbf{Y}^{\tau_2}_K\big]$, which can be reshaped into a tensor form $\underline{\Y}^{\tau_2}\in\C^{NU\times K\times\tau_2}$. We define the cascaded channel collecting all users as $\bar{\mathbf{Q}}\in\C^{NU\times \frac{M(\bar{M}+1)}{2}K}$, given by \begin{equation}
    \bar{\mathbf{Q}}=\big[\bar{\mathbf{Q}}_1,\bar{\mathbf{Q}}_2,\ldots,\bar{\mathbf{Q}}_K\big],
\end{equation} 
and can be reshaped into a tensor form $\underline{\bar{\mathbf{Q}}}\in\C^{NU\times K\times \frac{M(\bar{M}+1)}{2}}$. In Phase II, the objective is to recover the cascaded channel $\underline{\bar{\mathbf{Q}}}$ from the received pilot observations $\underline{\mathbf{Y}}^{\tau_2}$. By exploiting the correlations present across multiple dimensions of $\underline{\bar{\mathbf{Q}}}$, i.e., the $NU$ and $K$ dimensions, the pilot overhead required to estimate $\underline{\bar{\mathbf{Q}}}$ can be significantly reduced. In contrast, conventional LS-based estimation requires at least $\tau = \frac{M(\bar{M}+1)}{2}$ subframes to uniquely estimate $\underline{\bar{\mathbf{Q}}}$, such that the training scattering matrix $\tilde{\bm{\Phi}}^{\tau}$ has full row rank when estimating each $\bar{\mathbf{Q}}_k$. This leads to a total of $KU\frac{M(\bar{M}+1)}{2}$ pilot transmission time slots, which is prohibitively large. Therefore, we design the DACE to effectively exploit the multi-dimensional correlations in the cascaded channel $\underline{\bar{\mathbf{Q}}}$ to significantly reduce the pilot overhead.

\textit{Remark 2:} Unlike the training protocols in \cite{channel_estimation_hongyu,tensor_estiamte_ce_bd_ris,tensor_predict_estimate_ce_bd_ris}, which use all training slots for channel estimation with fixed channel-independent training patterns, our two-phase protocol allocates a small portion of the training slots to first optimize the training scattering
matrices. This protocol enables a channel-dependent design of $\tilde{\bm\Phi}^{\tau_2}$ based on the cascaded channel information embedded in $\underline{\Y}^{\tau_1}$. Moreover, the length of $\tilde{\bm\Phi}^{\tau_2}$, i.e., $\tau_2$, can be flexibly configured, rather than being fixed as in most channel-independent training designs.

\section{Training Scattering Matrix Learning and Dual-Attention Channel Estimation}\label{sec:J_TSML_CE_Network}

\begin{figure*}
    \centering
    \includegraphics[width=0.8\textwidth]{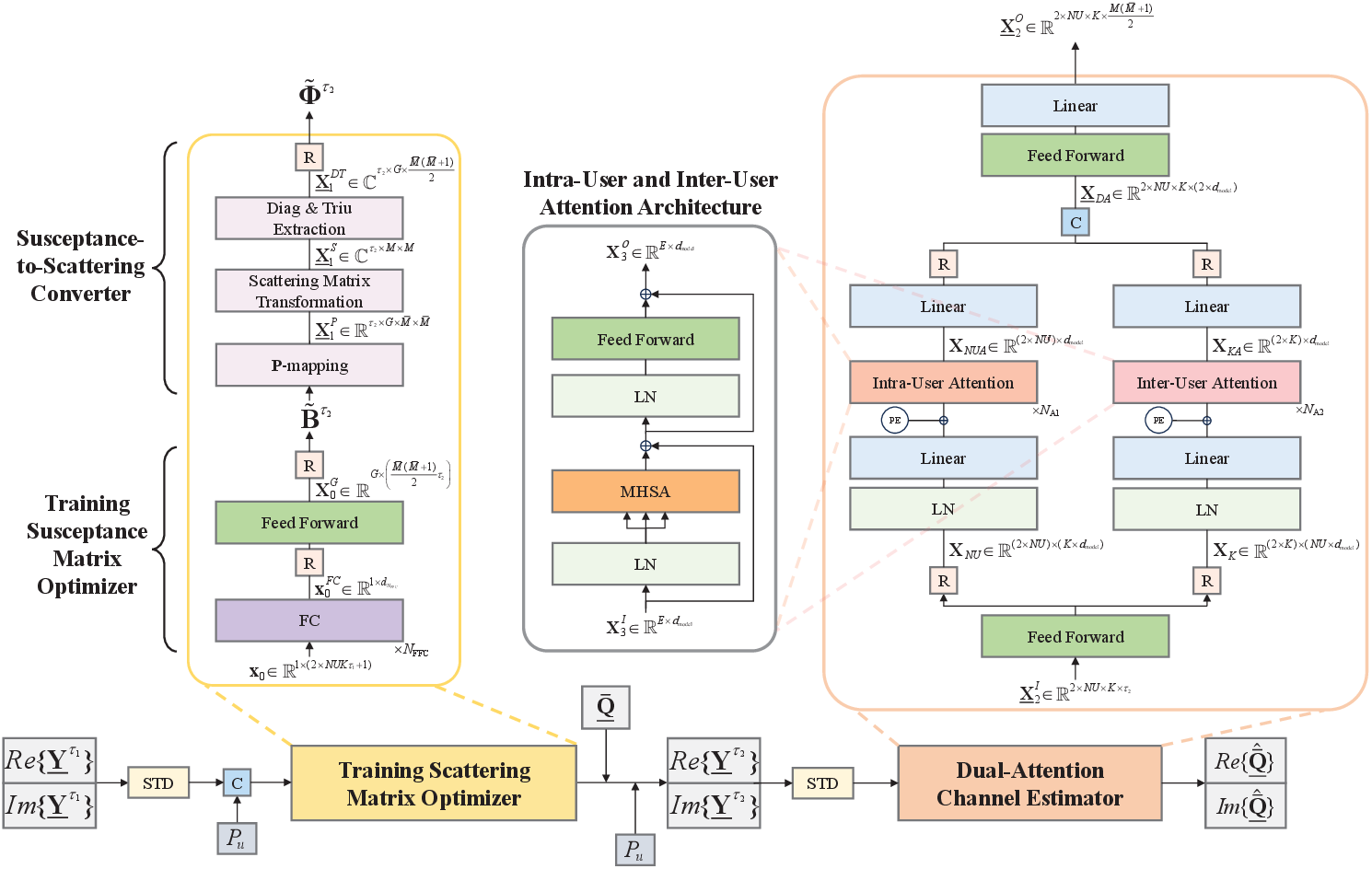}
    \caption{The overall architecture of the joint training scattering matrix learning and channel estimation framework (JTSMLCEF). MHSA: multi-head self-attention. FC: fully-connected layer. LN: layer normalization. STD: standardization. C: concatenation operation.  R: rearrangement operation. Diag \& Triu: diagonal and upper-triangular entries.}\label{fig:network}
\end{figure*}

In this section, we propose the TSMO and the DACE to enable adaptive training scattering matrix learning and cascaded channel estimation. The TSMO generates the training scattering matrix $\tilde{\bm{\Phi}}^{\tau_2}$ in Phase I uplink transmission, while the DACE learns the intra- and inter-user features of the multi-dimensional cascaded channel $\underline{\bar{\mathbf{Q}}}$ in Phase II based on the optimized training scattering matrix. Through this joint design, the TSMO and DACE collaboratively minimize the channel estimation error. The overall architecture of the TSMO and the DACE is shown in Fig. \ref{fig:network}. The uplink cascaded channel estimation problem can be formulated as\begin{subequations}\label{eq:min_Q_Q_hat}
    \begin{align}                    \min_{\bm\theta_{\mathsf{CE}},\tilde{\bm{\Phi}}^{\tau_2}}~& \mathcal{L}(\underline{\bar{\mathbf{Q}}}, \hat{\underline{\bar{\mathbf{Q}}}})\label{eq:Q_Q_hat_a}\\
    \mathrm{s.t.}~~~&\big[\Re\{\hat{\underline{\bar{\mathbf{Q}}}}\},\Im\{\hat{\underline{\bar{\mathbf{Q}}}}\}\big]= F_{\mathsf{CE}}\left(\big[\Re\{\underline{\Y}^{\tau_2}\},\Im\{\underline{\Y}^{\tau_2}\}\big];\bm\theta_{\mathsf{CE}}\right),\label{eq:Q_Q_hat_b}\\
    &\mathbf{Y}^{\tau_2}_k
    = \sqrt{P_u}\bar{\mathbf{Q}}_k\tilde{\bm{\Phi}}^{\tau_2} + \widetilde{\mathbf{N}}^{\tau_2},~\forall k\in\mathcal{K},\label{eq:Q_Q_hat_c}\\
    &\tilde{\bm\Phi}^{\tau_2}= F_{\mathsf{SMO}}\left(\big[\Re\{\underline{\Y}^{\tau_1}\},\Im\{\underline{\Y}^{\tau_1}\}\big];\bm\theta_{\mathsf{SMO}}\right),\label{eq:Q_Q_hat_d}
    \end{align}
\end{subequations}where $\hat{\underline{\bar{\mathbf{Q}}}}$ denotes the estimate of $\underline{\bar{\mathbf{Q}}}$, $F_{\mathsf{CE}}(\cdot;
\bm\theta_{\mathsf{CE}})$ and $F_{\mathsf{SMO}}(\cdot;
\bm\theta_{\mathsf{SMO}})$ are the DACE parameterized by $\bm\theta_{\mathsf{CE}}$ and the TSMO parameterized by $\bm\theta_{\mathsf{SMO}}$, respectively, and $\mathcal{L}$ is a loss function, e.g., the MSE loss function.

\subsection{Training Scattering Matrix Optimization}
Due to the specific BD-RIS constraints induced by the circuit topology of the reconfigurable impedance network i.e., unitarity and symmetry, the orthogonality-based scattering matrix design commonly used for D-RIS-aided systems is no longer applicable. Consequently, the scattering matrix must be carefully designed during uplink training to comply with the BD-RIS circuit constraints. This motivates the development of a dedicated training scattering matrix design. In this work, we propose to optimize $\tilde{\bm{\Phi}}^{\tau_2}$ with the objective of minimizing the channel estimation error in a data-driven manner. To guarantee that each scattering matrix $\bm\Phi^{t_2},~\forall t_2$, satisfies the unitary and symmetric constraints, the TSMO is decomposed into two components: a training susceptance matrix optimizer and a susceptance-to-scattering converter (SSC). 

\subsubsection{Training Susceptance Matrix Optimization}

Directly generating scattering coefficients using neural networks is infeasible due to the constraint imposed by a lossless BD-RIS, i.e., unitary scattering matrix. To address this issue, based on the microwave network theory \cite{pozar2009microwave}, we relate the scattering matrix\footnote{We use $\bm\Theta$ to denote the general BD-RIS scattering matrix, to distinguish it from $\bm\Phi$, which represents the uplink scattering matrix.} $\bm\Theta\in\C^{M\times M}$ of the BD-RIS to its admittance matrix $\tilde{\Y}\in\C^{M\times M}$ as\begin{equation}
    \bm\Theta=(\mathbf{I}_{M}+Z_0\tilde{\Y})^{-1}(\mathbf{I}_{M}-Z_0\tilde{\Y}),\label{eq:Theta_Y}
\end{equation}
where $Z_0$ denotes the characteristic impedance and is set to $50~\Omega$. Accordingly, for a lossless reciprocal BD-RIS, $\tilde{\Y}$ must be purely imaginary and can be expressed as $\tilde{\Y} = \jmath \mathbf{B}$, where $\mathbf{B} \in \mathbb{R}^{M \times M}$ denotes the susceptance matrix of the reconfigurable impedance network. This condition is equivalent to requiring the corresponding scattering matrix $\bm\Theta$ to be unitary. More specifically, we can relate each $\bm\Phi_{g}^{t_2}$ to $\mathbf{B}_{g}^{t_2},$ through\begin{equation}
    \bm\Phi_{g}^{t_2}=(\mathbf{I}_{\bar{M}}+Z_0\jmath\mathbf{B}_{g}^{t_2})^{-1}(\mathbf{I}_{\bar{M}}-Z_0\jmath\mathbf{B}_{g}^{t_2}),\label{eq:Phi_t2_g_B_t2_g}
\end{equation}
where $\mathbf{B}_{g}^{t_2}\in\mathbb{R}^{\bar{M}\times\bar{M}}$ denotes the susceptance matrix applied to the $g$-th group of BD-RIS in subframe $t_2,~\forall g\in\mathcal{G},~\forall t_2\in\mathcal{T}_2$. For a reciprocal BD-RIS, each $\mathbf{B}_{g}^{t_2}$ is also symmetric. Therefore, the same mapping matrix $\mathbf{P}$ can be used to extract the diagonal and upper-triangular entries of each $\mathbf{B}_{g}^{t_2}$, yielding $\mathbf{B}_{g}^{t_2}=\overline{\mathsf{vec}}(\mathbf{P}\bar{\mathbf{b}}_{g}^{t_2})$, where $\bar{\mathbf{b}}_{g}^{t_2}\in \mathbb{R}^{\frac{\bar{M}(\bar{M}+1)}{2} \times 1}$ collects the diagonal and upper-triangular entries of $\mathbf{B}_{g}^{t_2},~\forall g\in\mathcal{G}$. We further define $\bar{\mathbf{b}}^{t_2}=\big[(\bar{\mathbf{b}}_{1}^{t_2})^T,(\bar{\mathbf{b}}_{2}^{t_2})^T,\ldots,(\bar{\mathbf{b}}_{G}^{t_2})^T\big]^T\in\mathbb{R}^{\frac{M(\bar{M}+1)}{2}\times 1}$,
to collect all independent susceptance parameters of $\mathbf{B}^{t_2}$ in subframe $t_2,~\forall t_2\in\mathcal{T}_2$. Across $\tau_2$ subframes, we stack these vectors as\begin{equation}
    \tilde{\mathbf{B}}^{\tau_2}=\big[\bar{\mathbf{b}}^1,\bar{\mathbf{b}}^2,\ldots,\bar{\mathbf{b}}^{\tau_2}\big]\in\mathbb{R}^{\frac{M(\bar{M}+1)}{2}\times \tau_2},\label{eq:B_tilde_tau2}
\end{equation} 
which constitutes the training susceptance matrix for Phase II and can be uniquely transformed into the corresponding training scattering matrix $\tilde{\bm\Phi}^{\tau_2}$.

To stabilize neural-network training, the input received pilot signals $\underline{\mathbf{Y}}^{\tau_1}$ are first standardized by removing the mean and dividing by the standard deviation. To ensure fair evaluation, the mean and standard deviation are computed solely from the training dataset, and the same statistics are used to standardize the validation and test datasets. The uplink transmit power $P_u$ is assumed to be known at the BS, so it is incorporated as prior information during offline training and online deployment. The standardized $\underline{\mathbf{Y}}^{\tau_1}$ are then flattened and concatenated with $P_u$, and the resulting vector is fed into the TSMO. First, $N_{\mathsf{FFC}}$ fully-connected (FC) layers are employed to extract high-dimensional features from the input vector $\mathbf{x}_0\in\mathbb{R}^{1\times(2\times NUK\tau_1 +1)}$, the output vector $\mathbf{x}_0^{FC}\in\mathbb{R}^{1\times d_{N_{\mathsf{FFC}}}}$ can be expressed as\begin{equation}
    \mathbf{x}_0^{FC}=\text{ReLU}((\text{ReLU}(\mathbf{x}_0\mathbf{W}_1+\mathbf{b}_1)\cdots)\mathbf{W}_{N_\mathsf{FFC}}+\mathbf{b}_{N_\mathsf{FFC}}),
\end{equation}
where $\text{ReLU}$ denotes the rectified linear unit activation function, $\mathbf{W}_1\in\mathbb{R}^{(2\times NUK\tau_1 +1)\times d_1},\ldots,\mathbf{W}_{N_\mathsf{FFC}}\in\mathbb{R}^{d_{N_{\mathsf{FFC}}-1} \times d_{N_{\mathsf{FFC}}}}$ and $\mathbf{b}_1\in\mathbb{R}^{1\times d_1},\cdots,\mathbf{b}_{N_{\mathsf{FFC}}}\in\mathbb{R}^{1\times d_{N_{\mathsf{FFC}}}}$ denote the learnable weight matrices and bias vectors associated with the $N_{\mathsf{FFC}}$ FC layers, respectively. With the group-connected BD-RIS architecture, each RIS group carries the same type of local scattering information. To exploit this structural property, the global feature vector $\mathbf{x}_0^{FC}$ is reshaped into $G$ group-wise embeddings, denoted by $\mathbf{X}_0^{FCG}\in\mathbb{R}^{G\times \frac{d_{N_{\mathsf{FFC}}}}{G}}$, where each row corresponds to one BD-RIS group. A shared 2-layer feed-forward network, consisting of two linear layers with a ReLU activation in between, is then applied to each group embedding to regress the group-wise parameters $\mathbf{X}_0^{G}\in\mathbb{R}^{G\times \left(\frac{\bar{M}(\bar{M}+1)}{2}\tau_2\right)}$, which can be expressed as
\begin{equation}
  \mathbf{X}_0^{G}  =(\text{ReLU}(\mathbf{X}_0^{FCG}\mathbf{W}^{G}_1+\mathbf{b}^G_1))\mathbf{W}^G_2+\mathbf{b}^G_2,
\end{equation}
where $\mathbf{W}^{G}_1\in\mathbb{R}^{\frac{d_{N_{\mathsf{FFC}}}}{G}\times d_{G}}$ and $\mathbf{W}^{G}_2\in\mathbb{R}^{d_{G}\times \left(\frac{\bar{M}(\bar{M}+1)}{2}\tau_2\right)}$ are learnable weight matrices, and $\mathbf{b}^G_1\in\mathbb{R}^{1\times d_{G}}$ and $\mathbf{b}^G_2\in\mathbb{R}^{1\times \left(\frac{\bar{M}(\bar{M}+1)}{2}\tau_2\right)}$ are learnable bias vectors. By sharing the same transformation across all RIS groups, the proposed design enforces structural consistency and significantly reduces the number of learnable parameters, while preserving group-specific characteristics through the global feature representation $\mathbf{x}_0^{FC}$. Subsequently, the group-wise output $\mathbf{X}_0^{G}$ is reshaped to form the training susceptance matrix $\tilde{\mathbf{B}}^{\tau_2}$.

\subsubsection{Susceptance-to-Scattering Conversion}\label{sec:SSC}
The SSC is applied to transform the optimized $\tilde{\mathbf{B}}^{\tau_2}$ into the corresponding training scattering matrix $\tilde{\bm\Phi}^{\tau_2}$ while enforcing the physical constraints of a lossless reciprocal BD-RIS. This conversion is carried out through a sequence of reshaping and matrix operations. First, the input $\tilde{\mathbf{B}}^{\tau_2}$ is reshaped to recover the vectors $\bar{\mathbf{b}}_{g}^{t_2}$ for each group $g$ and each subframe $t_2$. For every $g$ and $t_2$, the $\mathbf{P}$ matrix is applied to map the diagonal and upper-triangular entries back to the corresponding susceptance matrix $\mathbf{B}_{g}^{t_2}$. The tensor stacking all $\mathbf{B}_{g}^{t_2},~\forall g\in\mathcal{G},~\forall t_2\in\mathcal{T}_2$, is denoted as $\underline{\mathbf{X}}_1^P\in\mathbb{R}^{\tau_2\times G\times \bar{M}\times \bar{M}}$. Every $\mathbf{B}^{t_2}$ can then be constructed from $[\underline{\mathbf{X}}_1^{P}]_{t_2,:,:,:}$. Next, all $\tau_2$ scattering matrices $\bm\Phi^{t_2}$ are computed in parallel via the transformation in \eqref{eq:Theta_Y}, resulting in the tensor $\underline{\mathbf{X}}_1^{S}\in\mathbb{C}^{\tau_2\times M\times M}$. Finally, the diagonal and upper-triangular entries of each $\bm\Phi_{g}^{t_2}$ are extracted from $\underline{\mathbf{X}}_1^{S}$, i.e., from the block $[\underline{\mathbf{X}}_1^S]_{t_2,(g-1)\bar{M}+1:g\bar{M},(g-1)\bar{M}+1:g\bar{M}}$. Collecting these entries over all groups and subframes forms the tensor $\underline{\mathbf{X}}_1^{DT}\in\mathbb{C}^{\tau_2\times G\times \frac{\bar{M}(\bar{M}+1)}{2}}$. After appropriate reshaping, the training scattering matrix $\tilde{\bm\Phi}^{\tau_2}$ is obtained.

\subsubsection{Online Control Signaling}
During online testing, only the training susceptance matrix optimizer is executed at the BS. Its output consists of the independent real-valued susceptance coefficients in $\tilde{\mathbf B}^{\tau_2}$. These coefficients are transmitted to the RIS controller instead of the complex-valued scattering matrices. The RIS controller maps the susceptance coefficients to the corresponding circuit-control parameters, such as the bias voltages of the tunable components. Once these parameters are configured, the reconfigurable impedance network realizes the associated scattering matrices. The SSC is employed during offline end-to-end training to enable differentiable optimization but does not need to be digitally evaluated in the online inference. The total number of coefficients generated for $\tau_2$ Phase-II subframes is $\frac{M(\bar M+1)}{2}\tau_2$. If each coefficient is represented using $e$ bits, the BS-to-RIS control payload has $e\frac{M(\bar M+1)}{2}\tau_2$ bits.

\subsection{Dual-Attention Channel Estimation}
With the optimized $\tilde{\bm\Phi}^{\tau_2}$, users transmit pilot signals over $\tau_2$ subframes, and the BS collects the corresponding received pilot signals $\underline{\mathbf{Y}}^{\tau_2}$. These signals are first standardized by removing the mean and dividing by the standard deviation\footnote{The same mean and standard deviation computed from the training dataset of $\underline{\mathbf{Y}}^{\tau_1}$ are used here.} and then fed into the DACE. Initially, a feed-forward embedding layer is applied to project the input tensor $\underline{\mathbf{X}}_2^I\in\mathbb{R}^{2\times NU\times K\times \tau_2}$ into a high-dimensional feature representation, denoted by $\underline{\mathbf{X}}_2^{EMB}\in\mathbb{R}^{2\times NU\times K\times d_{\mathsf{model}}}$. Specifically, the embedding is applied along the last dimension and shared across the remaining dimensions, i.e.,
\begin{equation}
    \underline{\mathbf{X}}_2^{EMB}  =(\text{ReLU}(\underline{\mathbf{X}}_2^I\mathbf{W}^{EMB}_1+\mathbf{b}^{EMB}_1))\mathbf{W}^{EMB}_2+\mathbf{b}^{EMB}_2,
\end{equation}
where $\mathbf{W}^{EMB}_1\in\mathbb{R}^{\tau_2\times d_{\mathsf{model}}}$ and $\mathbf{W}^{EMB}_2\in\mathbb{R}^{d_{\mathsf{model}}\times d_{\mathsf{model}}}$ are learnable weight matrices, $\mathbf{b}^{EMB}_1\in\mathbb{R}^{1\times 1\times 1\times d_{\mathsf{model}}}$ and $\mathbf{b}^{EMB}_2\in\mathbb{R}^{1\times 1\times 1\times d_{\mathsf{model}}}$ are learnable bias vectors broadcast along the $2\times NU\times K$ dimensions, and $d_{\mathsf{model}}$ denotes the representation dimension.

\subsubsection{Intra-User Correlation}\label{sec:intra-user_corr}
Within each cascaded channel $\bar{\mathbf{Q}}_k$, $k\in\mathcal{K}$, both rows and columns exhibit strong correlations among the channel coefficients. To reduce computational complexity, we focus on exploiting the correlations along the $NU$ dimension. These correlations arise from two main sources. First, spatial correlations are introduced between the signals received at different antennas, which typically depends on the antenna array geometry and the limited angular spread of the channel $\mathbf{H}_{IT}$. Second, structural correlations are induced by the Kronecker-product structure of the cascaded channel, where each coefficient of  $\mathbf{H}_{RI,k,g}$ scales the entire $\mathbf{H}_{IT,g},~\forall g\in\mathcal{G}$, resulting in repeated spatial patterns across different user antennas. The combination of these effects leads to highly correlated channel coefficients along the $NU$ dimension within $\bar{\mathbf{Q}}_k,~\forall k\in\mathcal{K}$.

\subsubsection{Inter-User Correlation}\label{sec:inter-user_corr}
In addition to the intra-user correlations, strong correlations also exist along the user dimension $K$ in the cascaded channel tensor $\underline{\bar{\mathbf{Q}}}$. These inter-user correlations mainly arise from two sources as well. First, users typically experience statistically similar wireless channels due to a shared propagation environment. Second, all cascaded channels $\bar{\mathbf{Q}}_k,~\forall k\in\mathcal{K}$, share the common RIS-BS channel $\mathbf{H}_{IT}$ in the Kronecker-product structure. Together, these effects result in pronounced correlations along the $K$ dimension of $\underline{\bar{\mathbf{Q}}}$.

\subsubsection{Intra-User and Inter-User Attention}
The self-attention mechanism has demonstrated strong capability in extracting complex dependencies across diverse domains, including sentences \cite{attention_is_all_you_need}, image patches \cite{attention_image_recognition}, and multi-path of wireless channels \cite{zhou_dual_attention}. By explicitly learning dependency relationships among input elements, self-attention is particularly well suited for capturing the multi-dimensional correlations inherent in $\underline{\bar{\mathbf{Q}}}$. As discussed in Sections~\ref{sec:intra-user_corr} and \ref{sec:inter-user_corr}, $\underline{\bar{\mathbf{Q}}}$ exhibits strong correlations along both the intra-user ($NU$) and inter-user ($K$) dimensions, which arise from different underlying mechanisms and are therefore heterogeneous. To effectively exploit these correlations, the DACE is constructed with two parallel MHSA branches, corresponding to the intra-user attention and inter-user attention, respectively. Compared with a cascaded structure of two MHSA mechanisms, the parallel design does not impose a predefined extraction order between these two heterogeneous correlations, and allows the intra-user and inter-user features to be extracted independently from the same embedded representation. By projecting the input features into multiple representation subspaces, MHSA enables the network to jointly capture diverse correlation patterns, leading to richer feature representations than single-head self-attention. Each MHSA module is integrated with layer normalization and residual connections to stabilize training and enable deep network connections, respectively, and combined with a feed-forward layer to enhance nonlinear feature representation. Together, these modules form the intra-user and inter-user attention, whose architectures are illustrated in Fig.~\ref{fig:network}. For clarity, we focus on elaborating the intra-user attention branch, as the inter-user attention branch follows the same design principle. In the intra-user attention branch, the embedded feature tensor $\underline{\mathbf{X}}_2^{EMB}$ is first reshaped into $\mathbf{X}_{NU}\in\mathbb{R}^{(2\times NU)\times (K\times d_{\mathsf{model}})}$, which concatenates the real and imaginary parts along the intra-user ($NU$) dimension, and then passes through a layer normalization operation. The output of the layer normalization, denoted by $\mathbf{X}_{NU}^{LN}\in\mathbb{R}^{(2\times NU)\times d^K_{\mathsf{model}}}$, can be formulated by
\begin{equation}
    \mathbf{X}_{NU}^{LN}=\frac{\mathbf{X}_{NU}-\mu_{\mathbf{X}_{NU}}}{\sigma_{\mathbf{X}_{NU}}} \odot \mathbf{g}_{NU}^{LN}+\mathbf{b}_{NU}^{LN},
\end{equation}
where $\mu_{\mathbf{X}_{NU}}=\frac{1}{d^K_{\mathsf{model}}} \sum_{i=1}^{d^K_{\mathsf{model}}}[\mathbf{X}_{NU}]_{:,i}\in\mathbb{R}^{(2\times NU)\times1}$ and $\sigma_{\mathbf{X}_{NU}}=\sqrt{\frac{1}{d^K_{\mathsf{model}}}\sum_{i=1}^{d^K_{\mathsf{model}}}\left([\mathbf{X}_{NU}]_{:,i} - \mu_{\mathbf{X}_{NU}}\right)^2}\in\mathbb{R}^{(2\times NU)\times1}$ denote the mean and standard deviation of $\mathbf{X}_{NU}$ computed along the feature dimension, respectively, $\mathbf{g}_{NU}^{LN}\in\mathbb{R}^{1\times d^K_{\mathsf{model}}}$ and $\mathbf{b}_{NU}^{LN}\in\mathbb{R}^{1\times d^K_{\mathsf{model}}}$ are learnable affine transformation parameters, and $d^K_{\mathsf{model}}=K\times d_{\mathsf{model}}$. Next, a linear layer is applied to reduce the feature dimension from $d_{\mathsf{model}}^K$ to $d_{\mathsf{model}}$, yielding $\mathbf{X}_{NU}^{L1}\in\mathbb{R}^{(2\times NU)\times d_{\mathsf{model}}}$. To incorporate positional information along the intra-user dimension, a sinusoidal positional encoding (PE) $\mathbf{P}_{NU}\in\mathbb{R}^{(2\times NU)\times d_{\mathsf{model}}}$ is added to $\mathbf{X}_{NU}^{L1}$. We adopt the sinusoidal PE scheme in \cite{attention_is_all_you_need}, defined as
\begin{align}
    [\mathbf{P}_{NU}]_{p,2j}&=\sin \left(\frac{p}{\xi^{2j/d_{\mathsf{model}}}}\right),\\
    [\mathbf{P}_{NU}]_{p,2j+1}&=\cos \left(\frac{p}{\xi^{2j/d_{\mathsf{model}}}}\right),
\end{align}
where $p\in[0,2\times NU-1]$ is the index along intra-user dimension, $j\in[0,d_{\mathsf{model}}/2-1]$ is the index along the $d_{\mathsf{model}}$ dimension, and $\xi$ is a hyper-parameter related to the value of $NU$. The input to the intra-user/inter-user attention\footnote{For the inter-user attention branch, the same processing pipeline is applied by exchanging $NU$ with the inter-user dimension $K$. Accordingly, we define $\mathbf{X}_{K}\in\mathbb{R}^{(2\times K)\times (NU\times d_{\mathsf{model}})}$, $\mathbf{X}_{K}^{LN}\in\mathbb{R}^{(2\times K)\times (NU\times d_{\mathsf{model}})}$, $\mathbf{X}_{K}^{L1}\in\mathbb{R}^{(2\times K)\times d_{\mathsf{model}}}$, and $\mathbf{P}_{K}\in\mathbb{R}^{(2\times K)\times d_{\mathsf{model}}}$ for inter-user attention, which correspond to $\mathbf{X}_{NU}$, $\mathbf{X}_{NU}^{LN}$, $\mathbf{X}_{NU}^{L1}$, and $\mathbf{P}_{NU}$, respectively.} is denoted by $\mathbf{X}_3^I\in\mathbb{R}^{E\times d_{\mathsf{model}}}$, and is given by\begin{equation}
\mathbf{X}_3^I=\mathbf{X}_{NU}^{L1}+\mathbf{P}_{NU} \quad \text{(intra-user attention)},
\end{equation}
or equivalently,
\begin{equation}
\mathbf{X}_3^I=\mathbf{X}_{K}^{L1}+\mathbf{P}_{K} \quad \text{(inter-user attention)},
\end{equation}
where $E\triangleq 2\times NU$ for the intra-user attention and $E\triangleq 2\times K$ for the inter-user attention. Within intra-user attention, the input $\mathbf{X}_3^I$ is first processed by a layer normalization operation, producing the output $\mathbf{X}_3^{LN1}\in\mathbb{R}^{E\times d_{\mathsf{model}}}$, given by
\begin{equation}
    \mathbf{X}_3^{LN1}=\frac{\mathbf{X}_3^I-\mu_{\mathbf{X}_3^I}}{\sigma_{\mathbf{X}_3^I}} \odot \mathbf{g}^{LN1}+\mathbf{b}^{LN1},
\end{equation}
where $\mu_{\mathbf{X}_3^I}\in\mathbb{R}^{E\times 1}$ and $\sigma_{\mathbf{X}_3^I}\in\mathbb{R}^{E\times 1}$ denote the mean and standard deviation of $\mathbf{X}_3^I$ computed along the feature dimension $d_{\mathsf{model}}$, respectively, $\mathbf{g}^{LN1}\in\mathbb{R}^{1\times d_{\mathsf{model}}}$ and $\mathbf{b}^{LN1}\in\mathbb{R}^{1\times d_{\mathsf{model}}}$ are learnable affine transformation parameters. For the MHSA, $\mathbf{X}_3^{LN1}$ is first projected into $N_{\mathsf{h}}$ sets of query, key and value matrices, denoted by $\mathbf{Q}^A_n\in\mathbb{R}^{E\times d_{\mathsf{k}}}$, $\mathbf{K}^A_n\in\mathbb{R}^{E\times d_{\mathsf{k}}}$, and $\mathbf{V}^A_n\in\mathbb{R}^{E\times d_{\mathsf{v}}}$, respectively, for each attention head $n\in\mathcal{N}=\{1,2,\ldots,N_{\mathsf{h}}\}$. Here, $d_{\mathsf{k}}=d_{\mathsf{v}}=d_{\mathsf{model}}/N_{\mathsf{h}}$. Specifically,
\begin{equation}
\mathbf{Q}^A_n=\mathbf{X}_3^{LN1}\mathbf{W}^Q_n,~\mathbf{K}^A_n=\mathbf{X}_3^{LN1}\mathbf{W}^K_n,~\mathbf{V}^A_n=\mathbf{X}_3^{LN1}\mathbf{W}^V_n,
\end{equation}
where $\mathbf{W}^Q_n\in\mathbb{R}^{d_{\mathsf{model}}\times d_{\mathsf{k}}}$, $\mathbf{W}^K_n\in\mathbb{R}^{d_{\mathsf{model}}\times d_{\mathsf{k}}}$, and $\mathbf{W}^V_n\in\mathbb{R}^{d_{\mathsf{model}}\times d_{\mathsf{v}}}$ are learnable weights. The output of the $n$-th self-attention head is computed via the scaled dot-product attention \cite{attention_is_all_you_need}, yielding $\mathbf{X}_{3,n}^{SA}\in\mathbb{R}^{E\times d_{\mathsf{v}}}$ as 
\begin{equation}
    \mathbf{X}_{3,n}^{SA}=\text{softmax}\left(\frac{\mathbf{Q}^A_n (\mathbf{K}_n^A)^T}{\sqrt{d_{\mathsf{k}}}}\right)\mathbf{V}^A_n,
\end{equation}
where $\text{softmax}(\cdot)$ is a row-wise softmax activation function. Finally, the outputs of all $N_{\mathsf{h}}$ attention heads are concatenated and projected to form the MHSA output $\mathbf{X}_{3}^{MHSA}\in\mathbb{R}^{E\times d_{\mathsf{model}}}$, given by \begin{equation}
    \mathbf{X}_{3}^{MHSA}=\big[\mathbf{X}_{3,1}^{SA},\mathbf{X}_{3,2}^{SA},\ldots,\mathbf{X}_{3,N_{\mathsf{h}}}^{SA}\big]\mathbf{W}^{O},
\end{equation}
where $\mathbf{W}^O\in\mathbb{R}^{d_{\mathsf{model}}\times d_{\mathsf{model}}}$ is a learnable weight matrix. Denote the outputs of the first residual connection, second layer normalization, and the feed-forward layer as $\mathbf{X}_{3}^{RC1}\in\mathbb{R}^{E\times d_{\mathsf{model}}}$, $\mathbf{X}_{3}^{LN2}\in\mathbb{R}^{E\times d_{\mathsf{model}}}$, and $\mathbf{X}_{3}^{FF}\in\mathbb{R}^{E\times d_{\mathsf{model}}}$, respectively. These operations are given by \begin{align}
    \mathbf{X}_{3}^{RC1} &= \mathbf{X}_{3}^{MHSA} + \mathbf{X}_{3}^{I},\\
    \mathbf{X}_{3}^{LN2} &= \frac{\mathbf{X}_{3}^{RC1}-\mu_{\mathbf{X}_{3}^{RC1}}}{\sigma_{\mathbf{X}_{3}^{RC1}}} \odot \mathbf{g}^{LN2}+\mathbf{b}^{LN2},
\end{align}
where $\mu_{\mathbf{X}_{3}^{RC1}}\in\mathbb{R}^{E\times 1}$ and $\sigma_{\mathbf{X}_{3}^{RC1}}\in\mathbb{R}^{E\times 1}$ denote the mean and standard deviation of $\mathbf{X}_{3}^{RC1}$ computed along $d_{\mathsf{model}}$ dimension, respectively, $\mathbf{g}^{LN2}\in\mathbb{R}^{1\times d_{\mathsf{model}}}$ and $\mathbf{b}^{LN2}\in\mathbb{R}^{1\times d_{\mathsf{model}}}$ are learnable affine transformation parameters. The output of the feed-forward layer is then given by
\begin{equation}
    \mathbf{X}_3^{FF}  =(\text{ReLU}(\mathbf{X}_3^{LN2}\mathbf{W}^{FF}_1+\mathbf{b}^{FF}_1))\mathbf{W}^{FF}_2+\mathbf{b}^{FF}_2,
\end{equation}
where $\mathbf{W}^{FF}_1\in\mathbb{R}^{d_{\mathsf{model}}\times d_{\mathsf{ff}}}$ and $\mathbf{W}^{FF}_2\in\mathbb{R}^{d_{\mathsf{ff}}\times d_{\mathsf{model}}}$ are learnable weight matrices, $\mathbf{b}^{FF}_1\in\mathbb{R}^{1\times d_{\mathsf{ff}}}$ and $\mathbf{b}^{FF}_2\in\mathbb{R}^{1\times d_{\mathsf{model}}}$ are learnable bias vectors. Finally, the output of the intra-user attention, denoted by $\mathbf{X}_3^O\in\mathbb{R}^{E\times d_{\mathsf{model}}}$, is obtained via a second residual connection as \begin{equation}
    \mathbf{X}_3^O = \mathbf{X}_3^{FF} + \mathbf{X}_{3}^{RC1}.
\end{equation}
In the proposed DACE, $N_{\mathsf{A1}}$ intra-user attention layers and $N_{\mathsf{A2}}$ inter-user attention layers are stacked. Let $\mathbf{X}_{NUA}\in\mathbb{R}^{(2\times NU)\times d_{\mathsf{model}}}$ and $\mathbf{X}_{KA}\in\mathbb{R}^{(2\times K)\times d_{\mathsf{model}}}$ denote the outputs of the stacked intra-user and inter-user attention layers, respectively. These outputs are then passed through two linear layers, yielding $\mathbf{X}^{L2}_{NUA}\in\mathbb{R}^{(2\times NU)\times (K\times d_{\mathsf{model}})}$ and $\mathbf{X}^{L2}_{KA}\in\mathbb{R}^{(2\times K)\times (NU\times d_{\mathsf{model}})}$, respectively. After reshaping, we obtain the tensors $\underline{\mathbf{X}}_{NUA}\in\mathbb{R}^{2\times NU\times K\times d_{\mathsf{model}}}$ and $\underline{\mathbf{X}}_{KA}\in\mathbb{R}^{2\times NU\times K\times d_{\mathsf{model}}}$, respectively. The two tensors are concatenated along the feature dimension to form the dual-attention tensor $\underline{\mathbf{X}}_{DA}\in\mathbb{R}^{2\times NU\times K\times (2\times d_{\mathsf{model}})}$. A feed-forward layer is then applied to merge the dual-attention features, denoted by $\underline{\mathbf{X}}_{DA}^{FF}\in\mathbb{R}^{2\times NU\times K\times d_{\mathsf{model}}}$, which can be expressed as \begin{equation}
    \underline{\mathbf{X}}_{DA}^{FF}  =(\text{ReLU}(\underline{\mathbf{X}}_{DA}\mathbf{W}^{DA}_1+\mathbf{b}^{DA}_1))\mathbf{W}^{DA}_2+\mathbf{b}^{DA}_2,
\end{equation}
where $\mathbf{W}^{DA}_1\in\mathbb{R}^{(2\times d_{\mathsf{model}})\times d_{\mathsf{model}}}$ and $\mathbf{W}^{DA}_2\in\mathbb{R}^{d_{\mathsf{model}}\times d_{\mathsf{model}}}$ are learnable weight matrices shared across the $2\times NU\times K$ dimensions, $\mathbf{b}^{DA}_1\in\mathbb{R}^{1\times 1\times 1\times d_{\mathsf{model}}}$ and $\mathbf{b}^{DA}_2\in\mathbb{R}^{1\times 1\times 1\times d_{\mathsf{model}}}$ are learnable bias vectors broadcast along the $2\times NU\times K$ dimensions. Finally, $\underline{\mathbf{X}}_{DA}^{FF}$ is fed into a linear layer to generate the DACE output $\underline{\mathbf{X}}_{2}^{O}\in\mathbb{R}^{2\times NU\times K\times \frac{M(\bar{M}+1)}{2}}$, which corresponds to the real and imaginary parts of the estimated cascaded channel tensor $\hat{\underline{\bar{\mathbf{Q}}}}$, i.e., $\Re\{\hat{\underline{\bar{\mathbf{Q}}}}\}=[\underline{\mathbf{X}}_{2}^{O}]_{1,:,:,:}$ and $\Im\{\hat{\underline{\bar{\mathbf{Q}}}}\}=[\underline{\mathbf{X}}_{2}^{O}]_{2,:,:,:}$.

\textit{Remark 3:} Existing BD-RIS channel estimation methods exploit channel dependencies through model-driven structures \cite{low_overhead_ce_bd_ris,tensor_estiamte_ce_bd_ris,tensor_predict_estimate_ce_bd_ris,semi_blind_tensor_ce_bd_ris,lowoverhead_bd_ris_group}. Specifically, \cite{low_overhead_ce_bd_ris,lowoverhead_bd_ris_group} exploit the channel dependency induced by the common RIS-BS channel, showing that the cascaded channel blocks associated with different users, user antennas, and BD-RIS elements are scaled versions of a common reference
cascaded channel. \cite{tensor_estiamte_ce_bd_ris,tensor_predict_estimate_ce_bd_ris,semi_blind_tensor_ce_bd_ris} exploit the multi-linear structure of BD-RIS-aided channels through tensor decomposition. However, these methods do not explicitly exploit the spatial correlations among antennas or the statistical correlations among RIS-user channels in multi-user systems. In contrast, the proposed DACE learns richer multi-dimensional channel dependencies, including Kronecker-product dependencies, spatial antenna correlations, and inter-user statistical correlations.

\subsection{Joint Learning}
The TSMO and the DACE are jointly trained in an end-to-end manner by minimizing the MSE between $\hat{\underline{\bar{\mathbf{Q}}}}$ and the corresponding label $\underline{\bar{\mathbf{Q}}}$, which is the loss function $\mathcal{L}$ in \eqref{eq:Q_Q_hat_a}, i.e., \begin{equation}
    \frac{1}{N_{\mathsf{bs}}} \sum\limits_{b=1}^{N_{\mathsf{bs}}}\frac{1}{N_{\mathsf{tot}}} \left\| \big[\Re\{\underline{\bar{\mathbf{Q}}}^b\},\Im\{\underline{\bar{\mathbf{Q}}}^b\}\big] - \big[\Re\{\hat{\underline{\bar{\mathbf{Q}}}}^b\},\Im\{\hat{\underline{\bar{\mathbf{Q}}}}^b\}\big] \right\|^2_F,\label{eq:mse_loss} 
\end{equation}
where $N_{\mathsf{bs}}$ is the batch size, $N_{\mathsf{tot}}=NUKM(\bar{M}+1)$ is the total number of real-valued coefficients in $\underline{\bar{\mathbf{Q}}}$, $\underline{\bar{\mathbf{Q}}}^b$ and $\hat{\underline{\bar{\mathbf{Q}}}}^b$ denote the $b$-th sample of $\underline{\bar{\mathbf{Q}}}$ and $\hat{\underline{\bar{\mathbf{Q}}}}$, respectively. It is worth noting that all operations involved in the SSC of the TSMO are differentiable. As a result, gradients of the MSE loss can be backpropagated through the DACE and the TSMO, enabling joint optimization of the training scattering matrix and the channel estimator. To improve numerical stability during training, the labels $\underline{\bar{\mathbf{Q}}}$ are normalized by the average cascaded channel gain computed from the training dataset.

\textit{Remark 4:} A few existing BD-RIS channel estimation works considered training scattering matrix designs \cite{channel_estimation_hongyu,low_overhead_ce_bd_ris,tensor_estiamte_ce_bd_ris,semi_blind_tensor_ce_bd_ris}, but these designs cannot replace the proposed TSMO. The reasons are twofold. First, apart from \cite{channel_estimation_hongyu}, which minimizes the MSE of the LS estimator, the designs in \cite{low_overhead_ce_bd_ris,tensor_estiamte_ce_bd_ris,semi_blind_tensor_ce_bd_ris} construct training scattering matrices to satisfy the rank, orthogonality, or identifiability conditions required by their estimators. Hence, these designs are estimator-specific and are not optimized for the channel estimation loss. Second, these designs do not explicitly enforce the reciprocity constraint on BD-RIS, and thus the generated training scattering matrices are not generally guaranteed to be symmetric.

\subsection{Extension to Other BD-RIS Architectures}
The JTSMLCEF can be easily extended to other BD-RIS architectures with architecture-dependent modifications. The key observation is that the two-phase training protocol and DACE are not tied to a specific circuit topology. Instead, the circuit topology of BD-RIS only affects its scattering matrix through the number and position of the independent tunable susceptance coefficients in the susceptance matrix. Therefore, extending JTSMLCEF to another BD-RIS architecture requires adapting the output dimension of training susceptance matrix optimizer and the architecture-dependent mapping in SSC. To make this clear, let $\mathcal{A}$ denote a specific BD-RIS architecture, and let $D_{\mathcal {A}}$ denote the number of independent susceptance coefficients of this architecture. For each $t_2$, the training susceptance matrix optimizer produces an architecture-dependent vector $\bar{\mathbf{b}}_{\mathcal{A}}^{t_2} \in \mathbb{R}^{D_{\mathcal{A}}\times 1}$. The susceptance matrix is constructed as $\mathsf{vec}(\mathbf{B}_{\mathcal{A}}^{t_2})
=\mathbf{T}_{\mathcal{A}}\bar{\mathbf{b}}_{\mathcal{A}}^{t_2}$, where $\mathbf{T}_{\mathcal{A}}\in
\{0,1\}^{M^2\times D_{\mathcal A}}$ is a global architecture-dependent mapping matrix that specifies the positions of independently tunable susceptance coefficients in the susceptance matrix according to the circuit topology. The corresponding scattering matrix is then obtained as
\begin{equation}
    \bm\Phi_{\mathcal{A}}^{t_2}
    =
    \left(\mathbf{I}_M+\jmath Z_0\mathbf{B}_{\mathcal{A}}^{t_2}\right)^{-1}
    \left(\mathbf{I}_M-\jmath Z_0\mathbf{B}_{\mathcal{A}}^{t_2}\right),~\forall t_2\in\mathcal{T}_2.
\end{equation}
Therefore, the SSC can be generalized to other BD-RIS architectures by replacing the $\mathbf{P}$-matrix mapping with $\mathbf{T}_{\mathcal{A}}$. For architectures with a block-diagonal scattering matrix, we have $D_{\mathcal A}=\frac{M(\bar{M}+1)}{2}$ for group-connected BD-RIS, $D_{\mathcal A}=2M-G$ for forest-connected BD-RIS, and $D_{\mathcal A}=Q_GM+M-G\dfrac{Q_G(Q_G+1)}{2}$ for cluster-connected BD-RIS, where $Q_G<\bar M$ denotes the stem width in each group for the cluster-connected BD-RIS \cite{general_architectures_xiaohua}. For architectures whose scattering matrix is full, the proposed framework can be applied by setting $G=1$. In this case, $D_{\mathcal{A}}=\frac{M(M+1)}{2}$ for fully-connected BD-RIS, $D_{\mathcal{A}}=2M-1$ for tree-connected BD-RIS, and $D_{\mathcal A}=QM+M-\dfrac{Q(Q+1)}{2}$ for band-/stem-connected BD-RIS, where $Q<M$ denotes the band width or stem width \cite{band_stem_zheyu,Q-stem_xiaohua}. In summary, extending JTSMLCEF to another BD-RIS architecture only requires adapting the output dimension of the training susceptance matrix optimizer, i.e., $D_{\mathcal{A}}\times \tau_2$, and replacing the group-wise $\mathbf{P}$-matrix mapping in the SSC with the global architecture-dependent mapping matrix $\mathbf{T}_{\mathcal{A}}$.

\section{Simulation Results}\label{sec:simulations}
In this section, we present simulation results to verify the effectiveness of the proposed JTSMLCEF for MU-MIMO BD-RIS-aided systems.

\subsection{Simulation Settings}
In this work, a geometry-based stochastic channel model, namely QuaDRiGa \cite{Quadriga_3D_channel_model}, is used to generate the simulation dataset, including both the RIS-BS and user-RIS channels. Two QuaDRiGa scenarios are considered: an indoor scenario based on the 3GPP TR 38.901 Indoor-Office model, and a UMi scenario based on the 3GPP TR 38.901 UMi model \cite{3gpp.38.901}. In both scenarios, the BS, BD-RIS, and users employ antenna arrays based on the 3GPP-3D antenna model specified in 3GPP TR 36.873 \cite{3gpp.36.873}. The BS is equipped with $N$ vertical antenna elements, the BD-RIS uses a rectangular array with $M$ elements, and each user employs $U$ vertical antenna elements.

\subsubsection{Setting of $\mathbf{H}_{RI}$}
\begin{figure}
    \centering
    \subfigure[Indoor Scenario]{
    \includegraphics[width=0.23\textwidth]{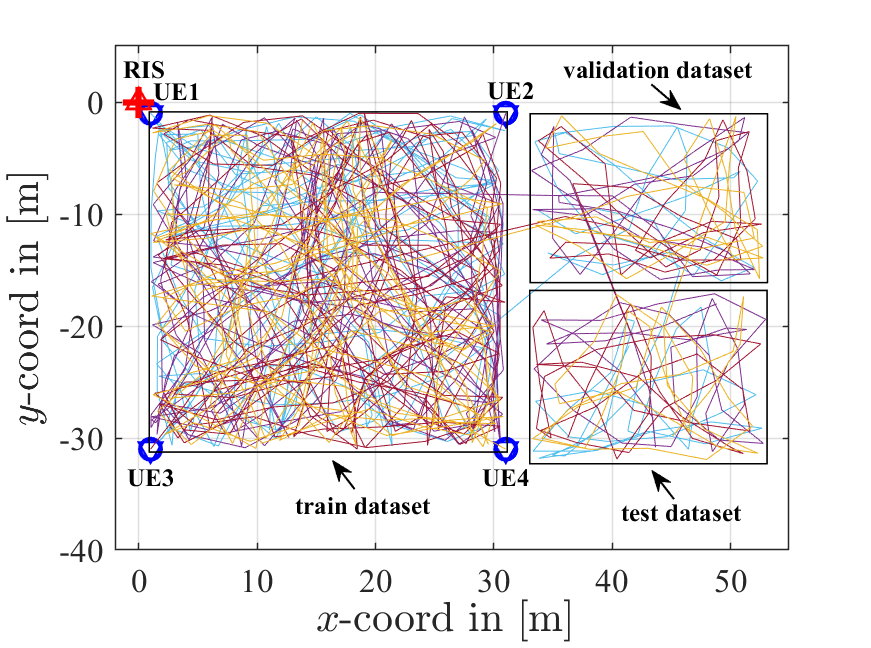}\label{fig:indoor_settings}}
    \subfigure[UMi Scenario]{
    \includegraphics[width=0.23\textwidth]{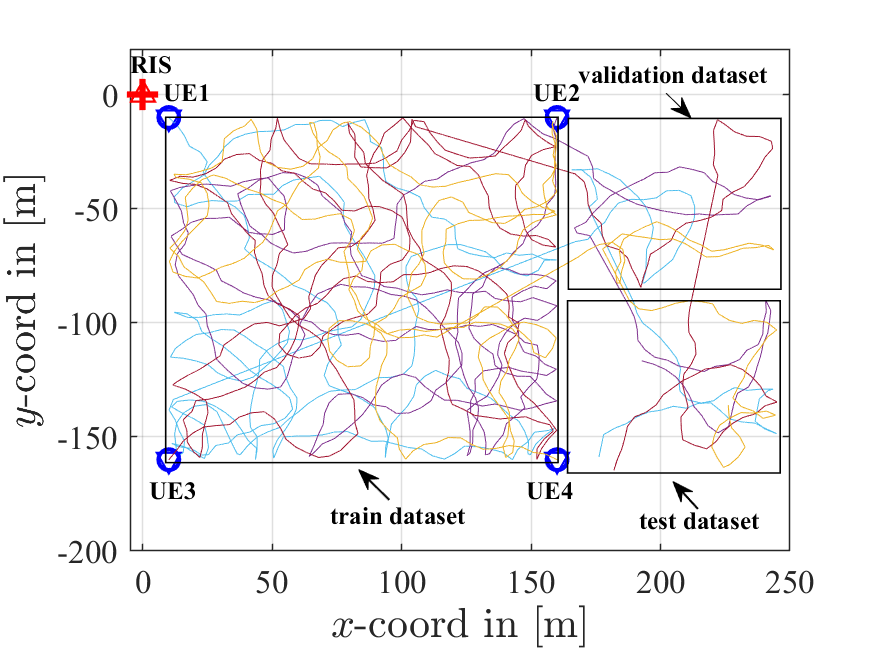}\label{fig:UMi_settings}}
    \caption{Layout of the RIS-to-user links and the trajectory of users under (a) indoor scenario, and (b) UMi scenario ($K=4$).}\label{fig:indoor_UMi_settings}
\end{figure}

The simulation layouts and user trajectories of the indoor and UMi scenarios are illustrated in Fig.~\ref{fig:indoor_settings} and Fig.~\ref{fig:UMi_settings}, respectively. In both scenarios, $K=4$ users move randomly within predefined areas.\footnote{For the indoor scenario, the area of training dataset is a $30\, \text{m} \times 30\,\text{m}$ rectangular region with coordinates $[1,31]\times[-31,-1]$ in the $(x,y)$ plane. The validation and test areas are defined as $[33,53]\times[-16,-1]$ and $[33,53]\times[-32,-17]$, respectively. For the UMi scenario, the area of training dataset is a $150\,\text{m} \times 150\,\text{m}$ rectangular region with coordinates $[10,160]\times[-160,-10]$. The validation and test areas are defined as $[165,245]\times[-85,-10]$ and $[165,245]\times[-165,-90]$, respectively.} To ensure data independence, three spatially disjoint areas are defined for training, validation, and testing. 

In both scenarios, each user moves along a trajectory of 1500 m to generate the training dataset and 150 m each for the validation and testing datasets, all at a constant speed of $1$ m/s. Channel samples are generated uniformly along each trajectory with a sampling rate of 80 samples per meter, corresponding to a sampling interval of 12.5 ms and approximately two samples per half-wavelength. For $\mathbf{H}_{RI,k},~\forall k\in\mathcal{K}$, each trajectory is divided into segments. For each segment, the propagation condition is randomly assigned to be either line-of-sight (LoS) or non-line-of-sight (NLoS) with equal probability, for both the indoor and UMi scenarios. A total of 100,000 samples are constructed for training, while 8,000 samples are constructed for each of the validation and testing datasets. The system parameters are summarized in Table~\ref{tab:system_setup}.

\subsubsection{Setting of $\mathbf{H}_{IT}$}

When generating the dataset of $\mathbf{H}_{IT}$, the BS and RIS are fixed at the locations specified in Table~\ref{tab:system_setup}. Since both the BS and RIS are static, the large-scale channel parameters, including the locations of scattering clusters, are generated once and shared across all channel samples. Different realizations of $\mathbf{H}_{IT}$ are obtained by varying the small-scale fading parameters across samples. For the indoor and UMi scenarios, the 3GPP TR 38.901 Indoor LoS and UMi LoS channel models are adopted, respectively, to generate $\mathbf{H}_{IT}$.

\begin{table}[htbp]
\centering
\caption{System parameter setup for indoor and UMi scenarios}
\label{tab:system_setup}
\resizebox{\columnwidth}{!}{%
\begin{tabular}{c|c|c}
\hline
\textbf{System Parameter} & \textbf{Indoor} & \textbf{UMi} \\ \hline
\textbf{Channel model parameter} & 3GPP TR 38.901 Indoor-Office & 3GPP TR 38.901 UMi \\ \hline
\textbf{BS location} & $(-10,-10,3)$ m & $(-100,-100,10)$ m \\ \hline
\textbf{RIS location} & $(0,0,3)$ m & $(0,0,10)$ m \\ \hline
\textbf{Height of all users} & $1$ m & $1.6$ m \\ \hline
\textbf{Segment length} & 5 m & 10 m \\ \hline
\textbf{Noise power} & -120 dBm & -140 dBm \\ \hline
\textbf{System frequency} & \multicolumn{2}{c}{6 GHz} \\ \hline
\textbf{Sampling interval} & \multicolumn{2}{c}{12.5 ms} \\ \hline
\textbf{$\mathbf{H}_{RI}$ per-segment LoS/NLoS condition} & \multicolumn{2}{c}{50\% LoS, 50\% NLoS} \\ \hline
\textbf{$\mathbf{H}_{IT}$ LoS/NLoS condition} & \multicolumn{2}{c}{LoS} \\ \hline
\textbf{Number of training samples} & \multicolumn{2}{c}{100,000} \\ \hline
\textbf{Number of validation samples} & \multicolumn{2}{c}{8,000} \\ \hline
\textbf{Number of testing samples} & \multicolumn{2}{c}{8,000} \\ \hline
\end{tabular}
}
\end{table}

\subsubsection{Experimental Settings}
The hyper-parameter settings for the JTSMLCEF are shown in Table~\ref{tab:hyper_parameter_settings}. The network is trained using Adam optimizer \cite{adam_optimizer} to minimize the loss function in \eqref{eq:mse_loss} until convergence, where the number of epochs is chosen empirically. The same number of attention heads $N_{\mathsf{h}}$ is used for both intra- and inter-user attention branches.

\begin{table}[htbp]
\centering
\caption{Hyper-parameter settings for the JTSMLCEF}
\label{tab:hyper_parameter_settings}
\scriptsize
\setlength{\tabcolsep}{3pt}      
\renewcommand{\arraystretch}{0.95} 
\begin{tabular}{c|c|c}
\hline
\textbf{Component} & \textbf{Hyper-parameter} & \textbf{Value} \\ \hline
\multirow{3}{*}{\textbf{TSMO}}
  & $N_{\mathsf{FFC}}$ & 3 \\ \cline{2-3}
  & $d_1,\ldots,d_{N_{\mathsf{FFC}}-1},d_{N_{\mathsf{FFC}}}$ & 400 \\ \cline{2-3}
  & $d_{G}$ & 400 \\ \hline

\multirow{6}{*}{\textbf{DACE}}
  & $d_{\mathsf{model}}$ & 256 \\ \cline{2-3}
  & $\xi$ & 1000 \\ \cline{2-3}
  & $N_{\mathsf{h}}$ & 2 \\ \cline{2-3}
  & $d_{\mathsf{ff}}$ & 512 \\ \cline{2-3}
  & $N_{\mathsf{A1}}$ & 3 \\ \cline{2-3}
  & $N_{\mathsf{A2}}$ & 3 \\ \hline

\multirow{2}{*}{\textbf{JTSMLCEF}}
  & $N_{\mathsf{bs}}$ & 400 \\ \cline{2-3}
  & initial learning rate & $1\times 10^{-4}$ \\ \hline

\end{tabular}
\end{table}

\subsubsection{Evaluation Metric, Baselines and $P_u$ Training Strategy} The NMSE is used as the metric to evaluate the channel estimation performance, which is defined as
\begin{equation}
    \text{NMSE} = \frac{1}{N_{\mathsf{test}}}\sum\limits_{j=1}^{N_{\mathsf{test}}} \frac{\left\|\underline{\bar{\mathbf{Q}}}^j - \hat{\underline{\bar{\mathbf{Q}}}}^j\right\|^2_F}{\big\|\underline{\bar{\mathbf{Q}}}^j \big\|^2_F},
\end{equation}
where $N_{\mathsf{test}}$ denotes the number of testing samples. We compare the proposed JTSMLCEF with two traditional channel estimation methods for BD-RIS, i.e., \cite{channel_estimation_hongyu} and \cite{low_overhead_ce_bd_ris}, which are referred to as \textbf{LS} and \textbf{LMMSE}, respectively. The \textbf{LS} method adopts the reduced-coefficient reformulation in this paper and uses the LS estimator \cite{channel_estimation_hongyu} to directly estimate $\bar{\mathbf{Q}}_k,~\forall k\in\mathcal{K}$, as in \eqref{eq:Y_k_tau1}. In \textbf{LMMSE} \cite{low_overhead_ce_bd_ris}, each group of BD-RIS is individually estimated using the linear minimum mean-squared error (LMMSE) estimator based on the low-overhead two-phase channel estimation scheme proposed for fully-connected BD-RIS in \cite{low_overhead_ce_bd_ris}. Note that since the scattering matrix designs proposed in \cite{channel_estimation_hongyu,low_overhead_ce_bd_ris} are asymmetric and therefore not applicable to reciprocal BD-RIS, the scattering matrices $\bm\Phi^t,~\forall t\in\mathcal{T}$, are randomly generated. In addition to BD-RIS channel estimation methods from the literature, we compare the JTSMLCEF with two deep learning-based baselines, namely \textbf{FCDNN} \cite{FC-DNN_baseline} and \textbf{GNN} \cite{learning_to_reflect_and_beamforming}. Specifically, \textbf{FCDNN} employs the FC-DNN-based channel estimation method \cite{FC-DNN_baseline}, and \textbf{GNN} adopts the graph neural network (GNN)-based channel estimation method proposed for D-RIS-aided systems in \cite{learning_to_reflect_and_beamforming}.\footnote{The FCDNN baseline employs four fully connected layers with a hidden dimension of 512, while the GNN baseline employs three message-passing layers with a hidden dimension of 256.} Furthermore, to evaluate the effectiveness of the TSMO, we remove it from JTSMLCEF and use randomly generated scattering matrices, resulting in the scheme denoted as \textbf{DACEN}. To evaluate the effectiveness of the DACE, we combine deep learning-based baselines with the TSMO to obtain another two schemes, denoted by \textbf{JTSML-FCDNN} and \textbf{JTSML-GNN}, which replace the DACE with an FC-DNN and the GNN proposed in \cite{learning_to_reflect_and_beamforming}, respectively. The hyper-parameters of FCDNN, GNN, DACEN, JTSML-FCDNN and JTSML-GNN are tuned to achieve their best performance for fair comparisons. Unless otherwise specified, e.g., in Fig. \ref{fig:Pu}, all schemes use the same total training duration of $\tau$ subframes. The baselines without Phase I use all subframes for their single-phase training.

In the simulations, the JTSMLCEF is trained using samples generated over a range of uplink transmit powers rather than a single fixed value. Specifically, for a target $P_u$, the framework is trained using samples with $P_u$ uniformly drawn from $[P_u - 2.5, P_u + 2.5]$ dBm. This training strategy allows a single pre-trained model to be directly deployed for any $P_u$ within the corresponding interval without retraining, improving robustness to moderate transmit power variations. For example, if the online deployment $P_u$ is $13.8$ dBm, the pre-trained model trained over the interval $[12.5, 17.5]$ dBm is directly used. All other deep learning-based schemes adopt the same training and deployment strategy as JTSMLCEF.

\subsection{Computational Complexity of JTSMLCEF}
During online deployment, JTSMLCEF executes the training susceptance matrix optimizer in TSMO followed by DACE. Let $d_{\mathsf{TSMO}}=d_1=\cdots=d_{N_{\mathsf{FFC}}}=d_G$. With $d_{\mathsf{ff}}=2d_{\mathsf{model}}$, and $\tau_2,NU,K,N_{\mathsf{A1}},N_{\mathsf{A2}}\ll d_{\mathsf{model}}$, the total online inference computational complexity is 
\begin{equation} 
\begin{aligned} \mathcal O(\mathsf{JTSMLCEF})&=\mathcal O\big(NUK\tau_1d_{\mathsf{TSMO}} +N_{\mathsf{FFC}}d_{\mathsf{TSMO}}^2\\
&+M(\bar M+1)\tau_2d_{\mathsf{TSMO}}+NUKd_{\mathsf{model}}^2\\ &+NUKM(\bar M+1)d_{\mathsf{model}}\big). \end{aligned} \end{equation} For $d_{\mathsf{TSMO}}=400$, $d_{\mathsf{model}}=256$, $d_{\mathsf{ff}}=512$, and three layers in each attention branch, JTSMLCEF requires $0.134$ GFLOPs and $6.555$ million parameters when $(M,N,K,U,\bar M,\tau_1,\tau_2)=(16,8,4,2,4,1,16)$. These values become $0.151$ GFLOPs and $7.471$ million parameters for $(64,8,4,2,8,1,64)$, respectively.

\subsection{Convergence Behavior of JTSMLCEF}

\begin{figure}
    \centering
    \subfigure[Indoor Scenario]{
    \includegraphics[width=0.23\textwidth]{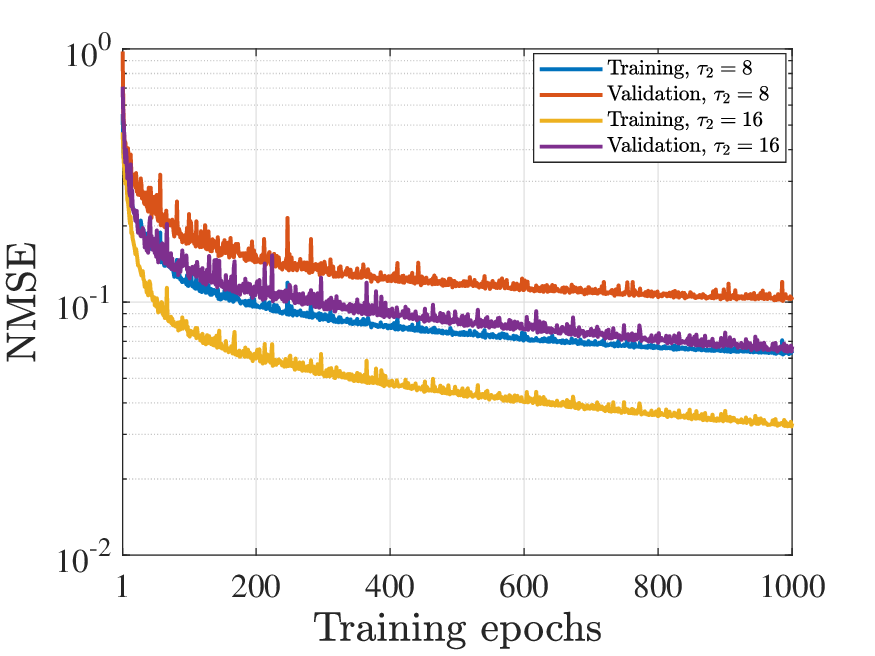}\label{fig:epoch_indoor}}
    \subfigure[UMi Scenario]{
    \includegraphics[width=0.23\textwidth]{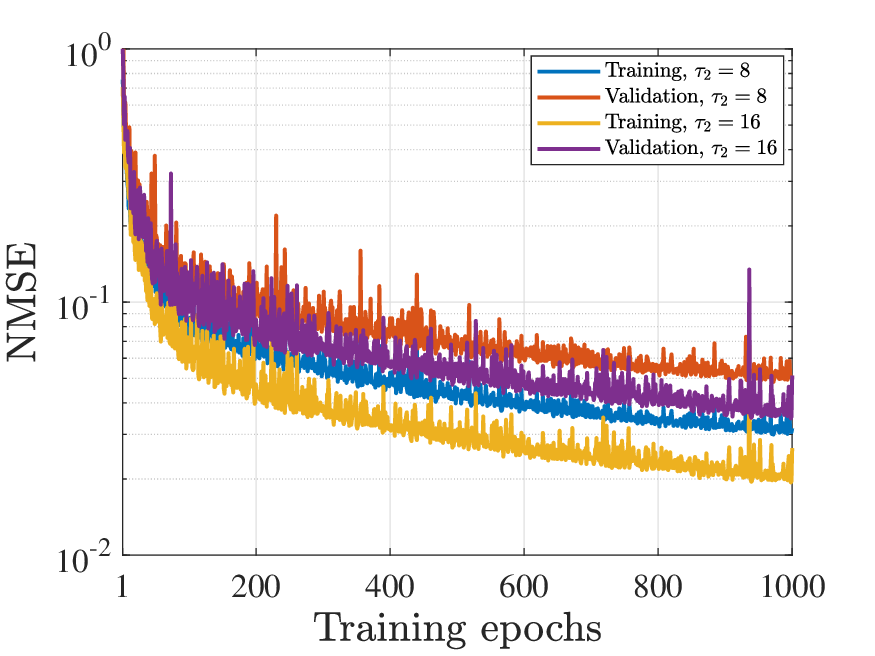}\label{fig:epoch_UMi}}
    \caption{Training and validation NMSEs versus the training epoch for $\tau_2=8$ and $\tau_2=16$: (a) indoor scenario with $P_u=20$ dBm; (b) UMi scenario with $P_u=40$ dBm ($M=16$, $\bar M=4$, $N=8$, $U=2$, $K=4$).}\label{fig:epoch}
\end{figure}

Although the joint learning problem is non-convex, Fig.~\ref{fig:epoch} demonstrates stable empirical convergence in both propagation scenarios. The validation NMSEs closely follow the decreasing trends of their corresponding training NMSEs in both scenarios. Although the NMSE curves exhibit occasional fluctuations, no persistent oscillation or divergence is observed. The UMi curves exhibit stronger epoch-to-epoch fluctuations than the indoor curves, which is consistent with greater sample-to-sample channel variability.

\subsection{Performance Evaluation of the Proposed JTSMLCEF}

\subsubsection{Impact of the subframe allocation}
\begin{figure}
    \centering
    \subfigure[Indoor Scenario]{
    \includegraphics[width=0.23\textwidth]{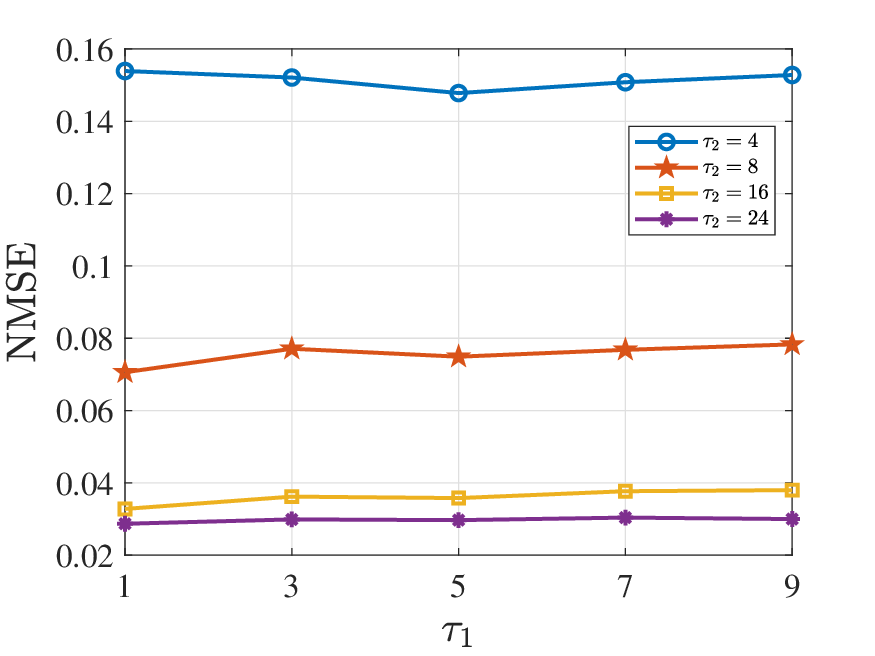}\label{fig:tau1_indoor}}
    \subfigure[UMi Scenario]{
    \includegraphics[width=0.23\textwidth]{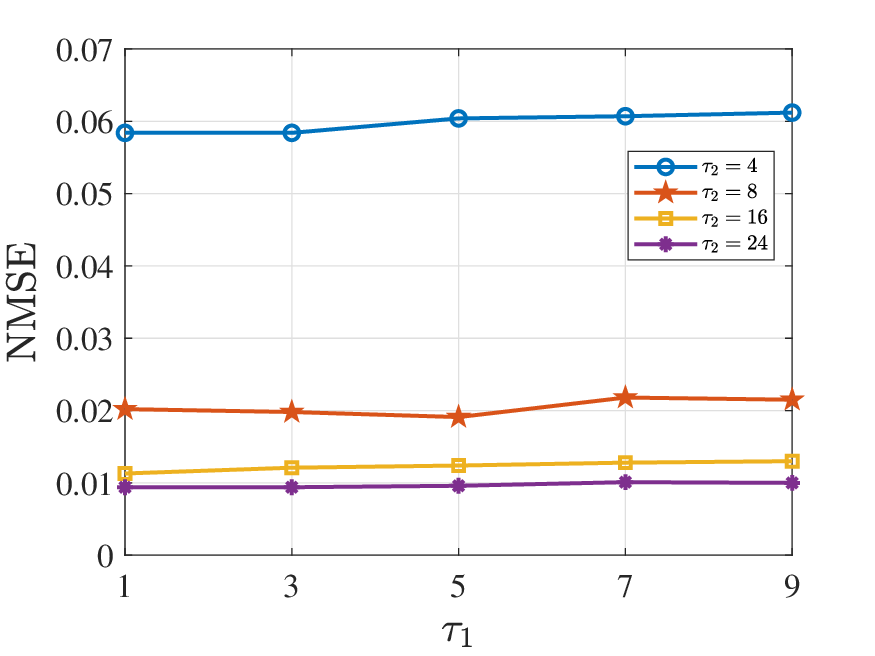}\label{fig:tau1_UMi}}
    \caption{NMSE performance versus different allocations of $\tau_1$ and $\tau_2$: (a) indoor scenario with $P_u=25$ dBm, and (b) UMi scenario with $P_u=40$ dBm ($M=16$, $N=8$, $K=4$, $U=2$, $\bar{M}=4$).}\label{fig:tau1}
\end{figure}

We first investigate the impact of the subframe allocation between Phase I and Phase II, i.e., $\tau_1$ and $\tau_2$, on the NMSE of the proposed JTSMLCEF. Figs.~\ref{fig:tau1_indoor} and \ref{fig:tau1_UMi} show NMSE versus $\tau_1$ for the indoor and UMi scenarios, respectively. In both scenarios, $\tau_1=1$ achieves the lowest or nearly the lowest NMSE, and increasing $\tau_1$ produces no systematic improvement.

To investigate the reason, we analyze $\tilde{\bm\Phi}^{\tau_2}$ generated over $8000$ UMi validation realizations under the same setting. Shuffling the Phase-I observations among realizations increases the NMSE by $8.6\%$ and $12.1\%$ for the models trained with $\tau_1=1$ and $\tau_1=9$, respectively, confirming that TSMO uses the available Phase-I information. For JTSMLCEF trained with $\tau_1=q$, we define the normalized information matrix as $\mathbf{G}(q)=
\frac{
\tilde{\boldsymbol{\Phi}}^{\tau_2}
(\tilde{\boldsymbol{\Phi}}^{\tau_2})^{H}
}{
\mathsf{Tr}\left(
\tilde{\boldsymbol{\Phi}}^{\tau_2}
(\tilde{\boldsymbol{\Phi}}^{\tau_2})^{H}
\right)
}$. The relative variations of $\mathbf G(q)$ across realizations are only $1.71\%$ and $3.62\%$ for $q=1$ and $q=9$, respectively, indicating a dominant common sensing geometry within each model. Let $\tilde{\bm{\Phi}}^{\tau_2}(q)=\mathbf{U}(q)\mathbf{R}(q)$ denote its reduced QR decomposition with $\mathbf U(q)\in\mathbb C^{\frac{M(\bar{M}+1)}{2}\times\tau_2}$. The normalized chordal
distance between the two sensing subspaces is  $d_{\mathrm c}(1,9)
=
\left(
1-
\frac{
\left\|
\mathbf U^H(1)\mathbf U(9)
\right\|_{F}^{2}
}{\tau_2}
\right)^{1/2}$. Averaging this distance over all validation realizations yields $0.796$. Thus, the two models learn substantially different $\tilde{\bm\Phi}^{\tau_2}$ despite achieving comparable NMSEs. This provides empirical evidence that the effective $\tilde{\bm\Phi}^{\tau_2}$ design is not unique: additional Phase-I information can change the learned design without improving the final NMSE. We therefore set $\tau_1=1$ in all subsequent simulations as the most overhead-efficient choice.

\subsubsection{Impact of uplink transmit power} 
\begin{figure}
    \centering
    \subfigure[Indoor Scenario]{
    \includegraphics[width=0.39\textwidth]{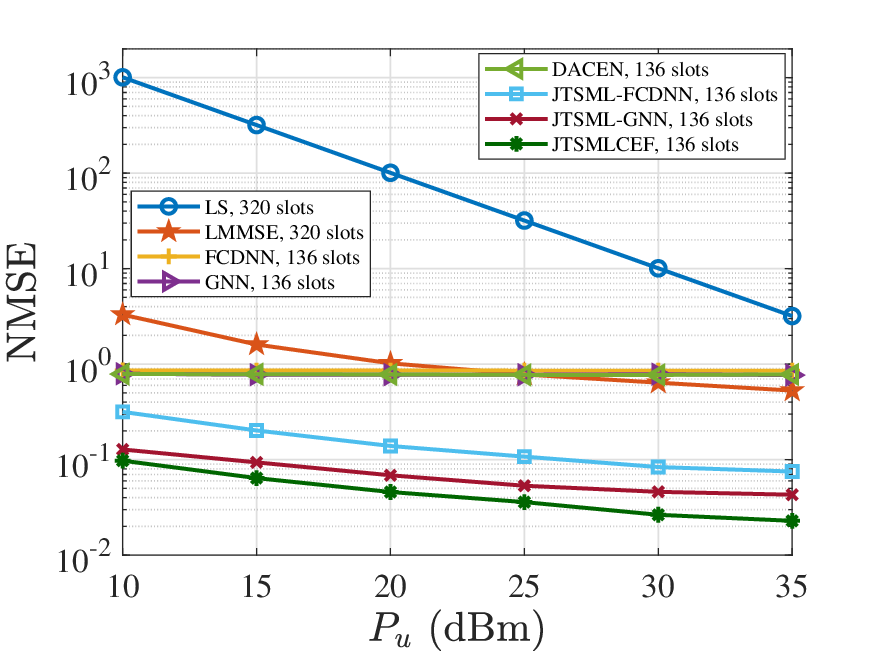}\label{fig:Pu_indoor}}
    \subfigure[UMi Scenario]{
    \includegraphics[width=0.39\textwidth]{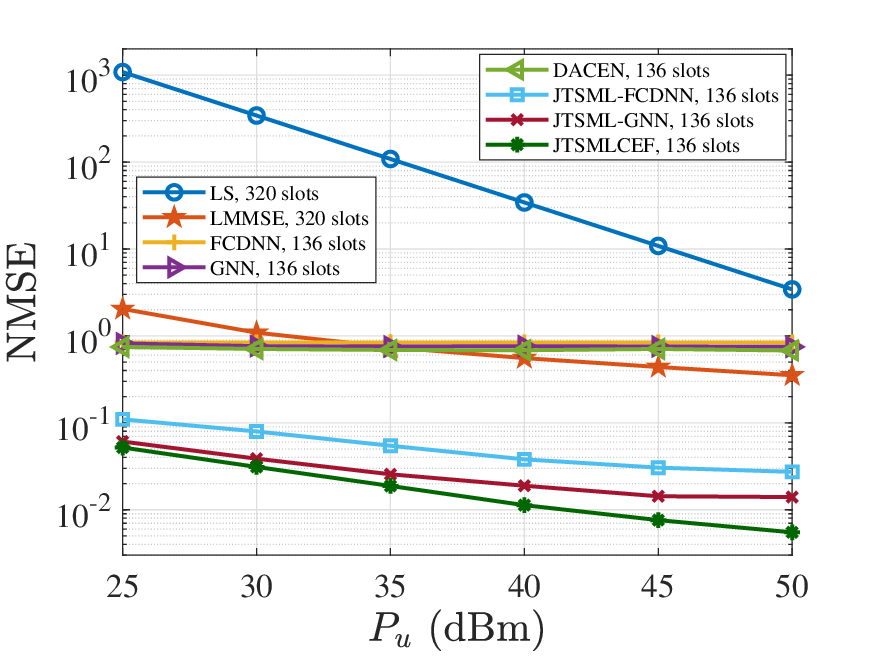}\label{fig:Pu_UMi}}
    \caption{NMSE performance versus the uplink transmit power $P_u$: (a) indoor scenario and (b) UMi scenario ($M=16$, $N=8$, $K=4$, $U=2$, $\bar{M}=4$).}\label{fig:Pu}
\end{figure}

Figs.~\ref{fig:Pu_indoor} and \ref{fig:Pu_UMi} show the NMSE performance versus $P_u$ in the indoor and UMi scenarios, respectively. The uplink transmit power $P_u$ is mapped to the average per-antenna signal-to-noise ratio (SNR), defined for user $k$ as $\text{SNR}_{k}=\frac{P_u \| \mathbf{H}_{IT}\bm\Phi\mathbf{H}_{RI,k}\|^2}{NU\sigma^2}$, and averaged over all users. The considered $P_u$ ranges ($10-35$ dBm and $25-50$ dBm, respectively) correspond to average SNR ranges of $4.4-29.4$ dB and $3.4-28.4$ dB for indoor and UMi scenarios, respectively. For the LS estimator, $\tau=\frac{M(\bar{M}+1)}{2}=40$ subframes are used to ensure a unique solution for each $\bar{\mathbf{Q}}_k$. For the LMMSE estimator, the pilot allocation between its two phases is optimized to achieve the best performance. To ensure a fair comparison between the LS and LMMSE estimators, both LS and LMMSE use a total of 320 pilot time slots, with $G=4$ and 80 time slots per group. In contrast, the deep learning-based methods use $\tau_2=16$, resulting in only 136 pilot time slots. We have the following key observations. \textit{First}, the proposed JTSMLCEF significantly outperforms LS, LMMSE, FCDNN and GNN baselines. Specifically, at $P_u = 25$ dBm and $40$ dBm for the indoor and UMi scenarios, respectively, JTSMLCEF achieves NMSE values of $0.0360$ and $0.0113$, reducing the NMSE achieved by the best-performing baselines (NMSE = $0.7767$ and $0.5556$) by $95.4\%$ and $98.0\%$. This performance gain arises mainly because, although the baseline methods exploit channel correlations either through explicit model-driven approaches or data-driven learning, they do not optimize the training scattering matrix under the reciprocal BD-RIS architecture. As a result, their NMSE performance is fundamentally limited, thereby highlighting the effectiveness of the proposed TSMO. \textit{Second}, JTSMLCEF further outperforms the JTSML-FCDNN and JTSML-GNN schemes. In the indoor and UMi scenarios at $P_u = 25$ dBm and $40$ dBm, respectively, the NMSE of JTSMLCEF is $33.4\%$ and $29.7\%$ of that of JTSML-FCDNN (NMSE = $0.1077$ and $0.0380$), and $67.4\%$ and $59.8\%$ of that of JTSML-GNN (NMSE = $0.0534$ and $0.0189$). These results demonstrate the effectiveness of the proposed DACE in capturing multi-dimensional correlations in multi-user cascaded channels, which are not effectively exploited by FC-DNN and GNN methods. \textit{Third}, all schemes (except LS) achieve lower NMSE in the UMi scenario than in the indoor scenario under comparable average SNR conditions. This performance gap is mainly attributed to the different channel statistics generated by QuaDRiGa. Indoor channels exhibit more severe small-scale fading and more frequent deep fades due to rich scattering and blockage under NLoS conditions, whereas UMi channels are statistically more stable and dominated by stronger geometric components, making them easier to estimate.

\subsubsection{Impact of RIS size and training length}
\begin{figure}
    \centering
    \subfigure[Indoor Scenario]{
    \includegraphics[width=0.39\textwidth]{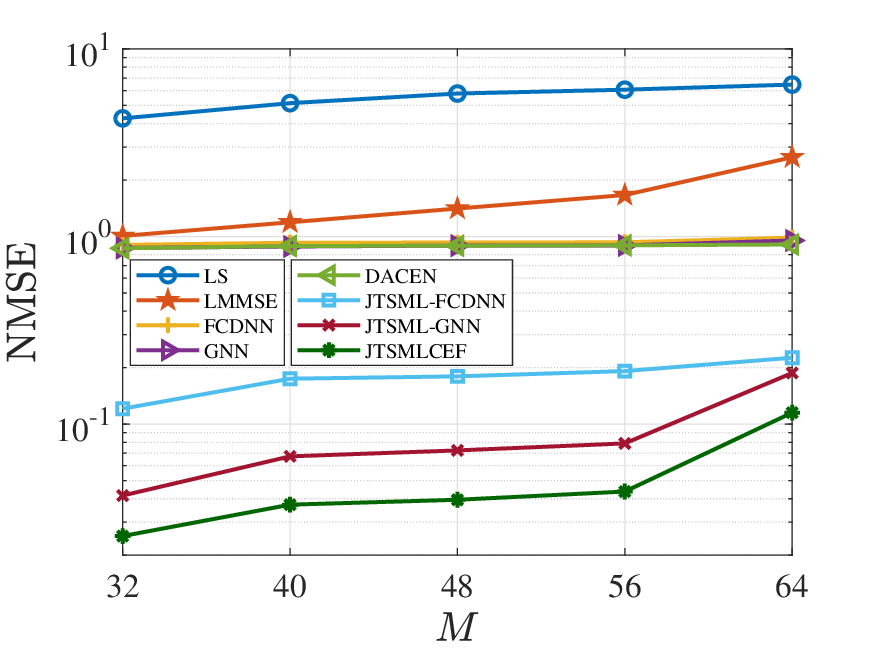}\label{fig:M_indoor}}
    \subfigure[UMi Scenario]{
    \includegraphics[width=0.39\textwidth]{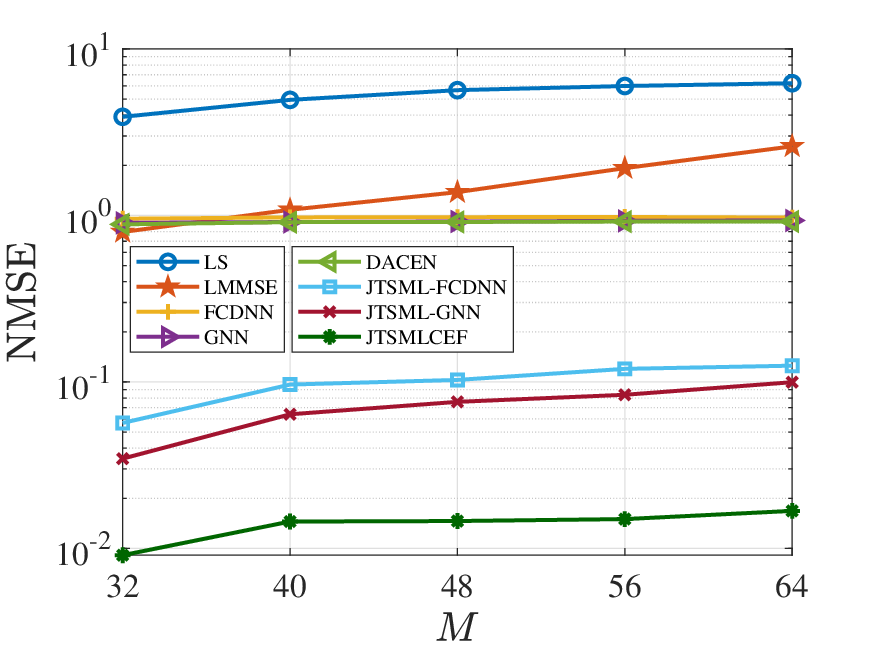}\label{fig:M_UMi}}
    \caption{NMSE performance versus the number of RIS elements $M$: (a) indoor scenario with $P_u=20$ dBm, and (b) UMi scenario with $P_u=40$ dBm ($N=8$, $K=4$, $U=2$, $\bar{M}=4$, $\tau=60$, $\tau_2=59$).}\label{fig:M}
\end{figure}

\begin{figure}
    \centering
    \subfigure[Indoor Scenario]{
    \includegraphics[width=0.39\textwidth]{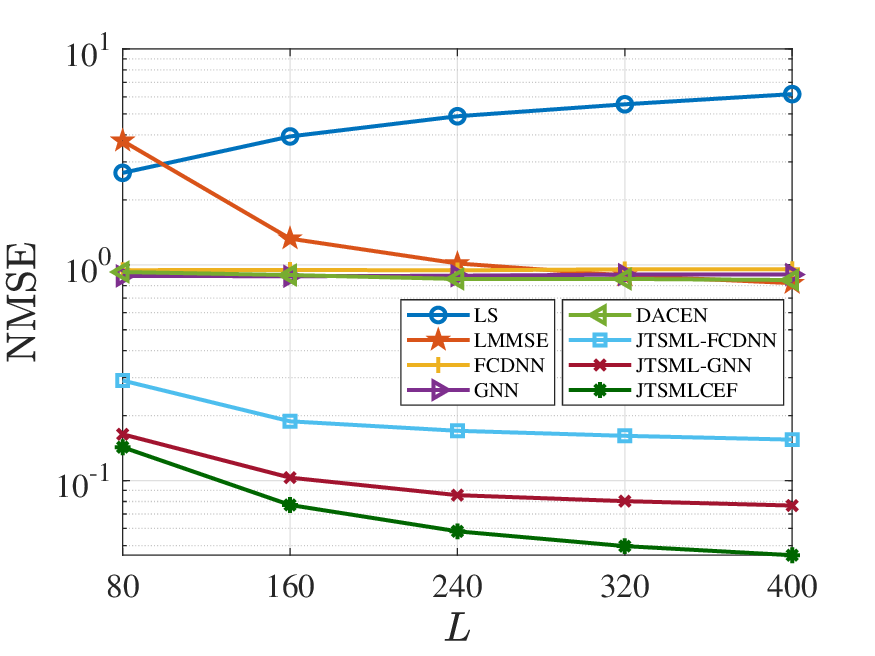}\label{fig:L_indoor}}
    \subfigure[UMi Scenario]{
    \includegraphics[width=0.39\textwidth]{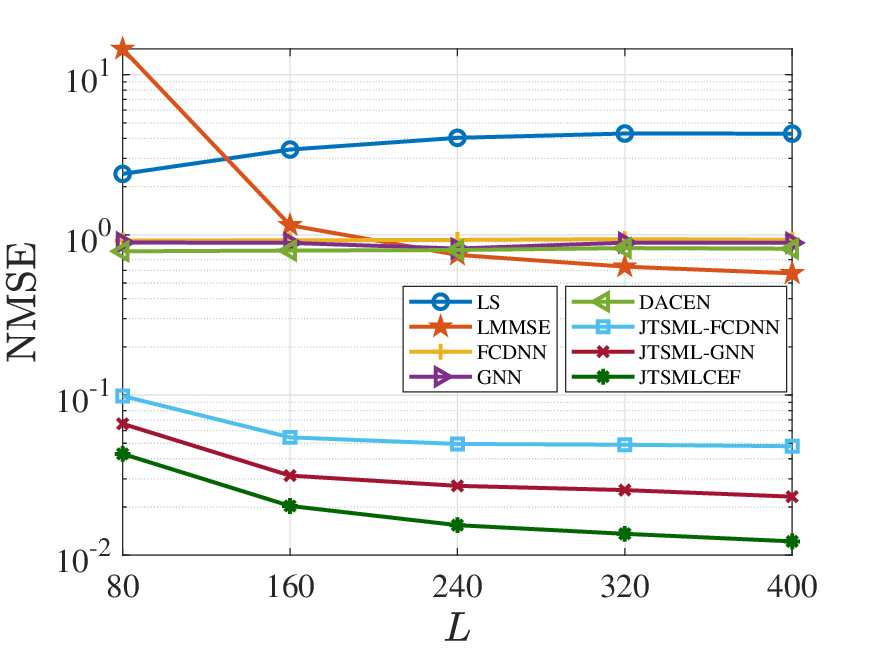}\label{fig:L_UMi}}
    \caption{NMSE performance versus training time slots $L$: (a) indoor scenario with $P_u=20$ dBm, and (b) UMi scenario with $P_u=40$ dBm ($M=16$, $N=8$, $K=4$, $U=2$, $\bar{M}=8$).}\label{fig:L}
\end{figure}

In Fig.~\ref{fig:M}, we fix the total number of training time slots to $L=480$ and increase the number of RIS elements $M$ in both indoor and UMi scenarios. JTSMLCEF consistently achieves the lowest NMSE as $M$ increases, while LS performs worst
because the available pilot length is insufficient for unique recovery. In Fig.~\ref{fig:L}, we investigate the impact of the total training time slots $L$ on the NMSE performance in both indoor and UMi scenarios. Using only $L=80$ pilots, JTSMLCEF achieves NMSEs of $0.1428$ and $0.0429$ for the indoor and UMi scenarios, respectively, whereas the best-performing traditional and deep learning-based baseline achieve NMSEs of $0.8222$ and $0.5744$ with $L=400$. As a result, compared with the best-performing baseline, the JTSMLCEF reduces the pilot overhead by $80\%$ while further reducing the NMSE by $82.6\%$ and $92.5\%$ in the indoor and UMi scenarios, respectively. The JTSML-FCDNN and JTSML-GNN methods show limited performance improvement once $L$ exceeds 160, indicating that they are unable to effectively exploit additional pilot observations to further enhance channel estimation accuracy.

\subsubsection{Ablation study of DACE}
\begin{figure}
    \centering
    \includegraphics[width=0.39\textwidth]{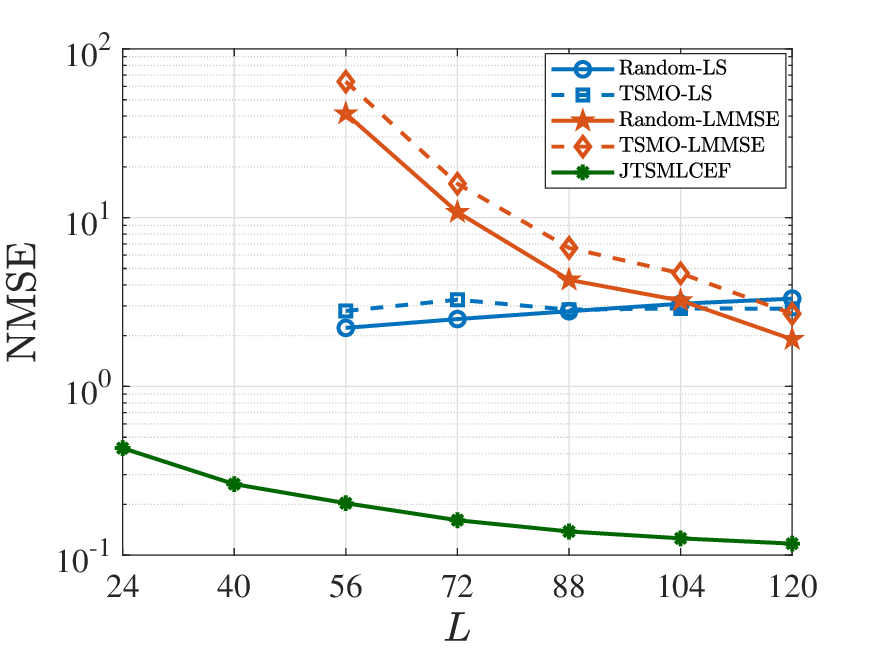}
    \caption{NMSE performance versus the training time slots $L$ in indoor scenario ($M=16$, $N=8$, $K=4$, $U=2$, $\bar{M}=8$, $P_u=20$ dBm). The LS- and LMMSE-based schemes are shown only for $L\geq56$, since the smaller values of $L$ do not satisfy the minimum training-slot requirement of the LMMSE estimator in \cite{low_overhead_ce_bd_ris}.}\label{fig:L_random_tsmo}
\end{figure}

Fig.~\ref{fig:L_random_tsmo} presents an ablation study to further validate the effectiveness of DACE. Random-LS and Random-LMMSE use randomly generated symmetric-unitary scattering matrices, whereas TSMO-LS and TSMO-LMMSE use TSMO-generated matrices. They all retain the LS estimator in \cite{channel_estimation_hongyu} and the LMMSE estimator in \cite{low_overhead_ce_bd_ris}, respectively. TSMO-LS provides no consistent improvement over Random-LS because the LS problem remains underdetermined, while TSMO-LMMSE consistently performs worse than Random-LMMSE. This is because TSMO is jointly optimized with DACE to generate informative pilot features, rather than to optimize the sensing-matrix conditioning required by LS or LMMSE. To verify this explanation, we evaluate the average minimum singular values of the Phase-I LMMSE sensing matrix in \cite{low_overhead_ce_bd_ris}; for Random-LMMSE, these values are $[0.0626,0.2121,0.3460,0.5065,0.6151]$, and for TSMO-LMMSE, they are $[0.0554,0.1156,0.2861,0.4544,0.5673]$. The consistently smaller values of TSMO-LMMSE indicate greater sensitivity to noise. These results show that the TSMO-generated matrices are estimator-dependent and do not necessarily improve traditional LS or LMMSE estimators. 

\subsubsection{Robustness to dynamic user scheduling}
\begin{figure}
    \centering
    \includegraphics[width=0.39\textwidth]{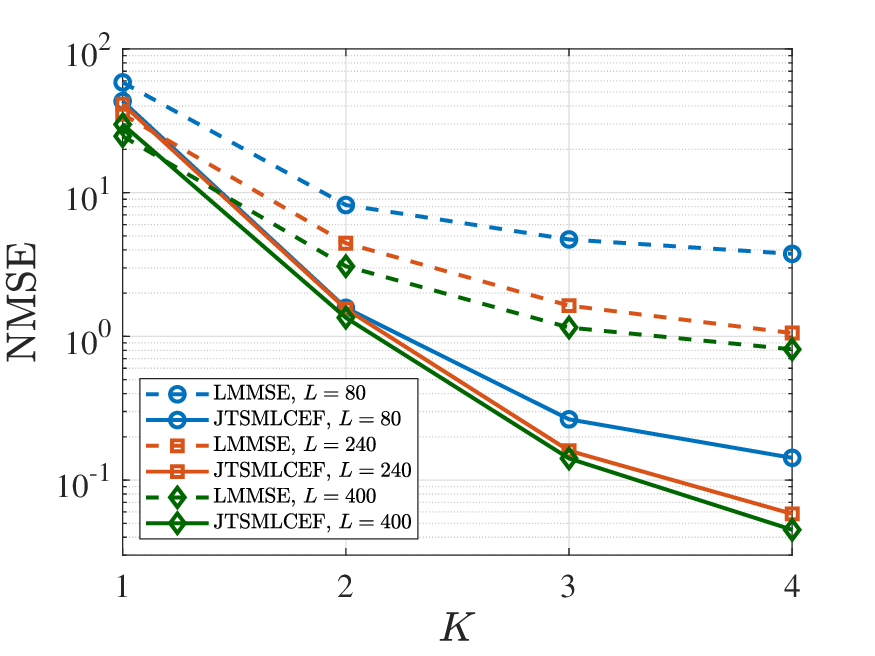}
    \caption{NMSE versus number of scheduled users $K$ under different training lengths $L$ ($M=16$, $N=8$, $U=2$, $\bar{M}=8$, $P_u=20$ dBm). }\label{fig:K}
\end{figure}

Fig.~\ref{fig:K} evaluates the robustness of JTSMLCEF to a dynamically changing number of scheduled users. For each training length $L$, the JTSMLCEF is trained only with the maximum supported number $K_{\max}=4$ and directly applied to $K\in\{1,2,3,4\}$ without retraining.\footnote{In practice, $K_{\max}$ can be selected according to the maximum multi-user scheduling capacity of the BS.} When $K<K_{\max}$, the received Phase-I and Phase-II pilot observations are cyclically duplicated to fill the $K_{\max}$ input positions, and the DACE outputs associated with the same active scheduled user are averaged. These results demonstrate that one model trained with $K_{\max}$ can accommodate moderate variations in the scheduled user load without requiring a separate model for each $K$; a separate single-user model may optionally be employed as a specialized operating mode.

\subsection{Quantitative Analysis of $T_{\mathsf{ctrl}}$}\label{sec:control_latency}
The online inference latency of the training susceptance matrix optimizer was measured on an NVIDIA GeForce RTX 5090 GPU. The average latency over $10{,}000$ inference runs was $0.114~\mathrm{ms}$ for $M=16$, $\bar M=4$, and $\tau_2=16$; and $0.115~\mathrm{ms}$ for $M=64$, $\bar M=8$, and $\tau_2=64$. Therefore, we use $T_{\mathsf{inf}}=0.115~\mathrm{ms}$ in the following evaluation.

\begin{figure}
\centering
\includegraphics[width=0.39\textwidth]{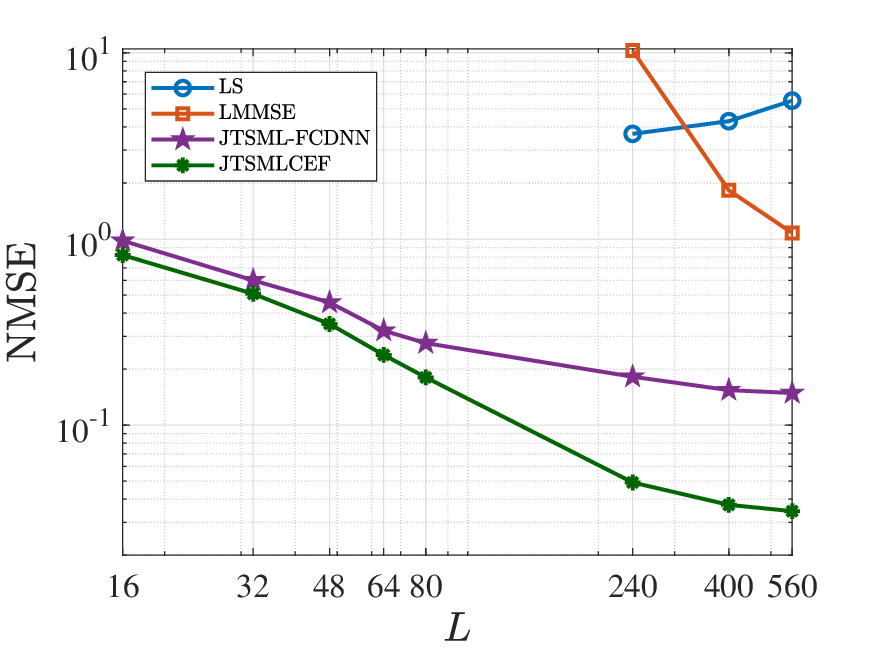}
\caption{NMSE versus the total number of training slots $L$ in the UMi scenario ($M=64$, $\bar M=8$, $N=8$, $K=4$, and $U=2$). The minimum training-slot requirement of the LMMSE scheme \cite{low_overhead_ce_bd_ris} is $L=184$ in this setting.}
\label{fig:L_low_delay}
\end{figure}

Fig.~\ref{fig:L_low_delay} shows that JTSMLCEF can operate with substantially shorter training lengths than the LS and LMMSE estimators. For $L=64$, we have $\tau_2=7$. Therefore, the number of susceptance coefficients generated by TSMO is $\frac{64(8+1)}{2}\times7=2016$. Using a 32-bit representation and a dedicated 1000BASE-LX Ethernet link with $1$-Gbit/s \cite{IEEE_Ethernet_standards}, the control link transmission delay is $T_{\mathsf{link}}=\frac{2016\times32}{10^9}=0.0645~\mathrm{ms}$. Consequently, $T_{\mathsf{ctrl}}=0.115+0.0645=0.180~\mathrm{ms}$. Under the $15$-kHz 5G NR numerology, a $1$-ms NR slot contains 14 OFDM symbols under the normal cyclic prefix \cite{3gpp.38.211}. Since one training time slot in our signal model corresponds to one OFDM pilot-symbol interval, $T_{\mathsf{slot}}=0.0714~\mathrm{ms}$. Thus, $T_{\mathsf{ctrl}}=0.180~\mathrm{ms}$ is equivalent to approximately $2.52$ training slot intervals, which is considered negligible.

\section{Conclusion}\label{sec:conclusion}
In this paper, we propose the JTSMLCEF, a learning-based cascaded channel estimation framework for BD-RIS-aided MU-MIMO systems, which follows a two-phase channel estimation protocol to jointly optimize the BD-RIS training scattering matrix with the TSMO and estimate the cascaded channels with the DACE in an end-to-end manner. 

Simulation results validate the effectiveness of the proposed JTSMLCEF. Specifically, the JTSMLCEF achieves the lowest NMSE compared to existing traditional and deep learning-based baselines, across various uplink transmit power levels, number of RIS elements and training time slots. The ablation and dynamic-user results further confirm the effectiveness of TSMO and DACE, and show that a single trained model can accommodate varying numbers of scheduled users. Overall, JTSMLCEF provides accurate channel estimation with low pilot overhead.

\bibliographystyle{IEEEtran}
\bibliography{refs}

\end{document}